\documentclass[%
 reprint,
 amsmath,amssymb,
 aps,
]{revtex4-2}

\usepackage{xcolor}
\usepackage{graphicx}
\usepackage{dcolumn}
\usepackage{bm}
\usepackage{soul}
\newcommand{\im}{{\rm i}}
\usepackage{float}
\usepackage{hyperref}

\hypersetup{
    colorlinks=true,
    linkcolor=blue,
    filecolor=black,
    anchorcolor=black,
    citecolor=blue,
    filecolor=black,
    urlcolor=blue
    }

\begin{document}

\preprint{APS/123-QED}

\title{Periodic skyrmionic textures via conformal cartographic projections
}%
\author{David Marco\textsuperscript{1, 2}}
\author{Isael Herrera\textsuperscript{1}}
\author{Sophie Brasselet\textsuperscript{1}}
\author{Miguel A. Alonso\textsuperscript{1, 3, 4, 5,}}
 \email{miguel.alonso@fresnel.fr}
\affiliation{%
  \textsuperscript{1}Aix Marseille Univ., CNRS, Centrale Marseille, Institut Fresnel, UMR 7249, 13397 Marseille Cedex 20, France\\
 \textsuperscript{2}Instituto de Bioingeniería, Universidad Miguel Hernández de Elche, 03202 Elche, Spain\\
 \textsuperscript{3}The Institute of Optics, University of Rochester, Rochester, NY14627, USA\\
 \textsuperscript{4}Center for Coherence and Quantum Optics, University of Rochester, Rochester, NY14627, USA\\
 \textsuperscript{5}Laboratory for Laser Energetics, University of Rochester, Rochester, NY14627, USA
}%
\date{\today}

\begin{abstract}
We find periodic skyrmionic textures via conformal cartographic projections that map either an entire spherical parameter space or a hemisphere onto every regular polygon that provides regular tessellations of the plane. These textures minimize the energy inherent to the mapping and preserve the sign of the Skyrme density throughout the entire space. We show that 2D spinor fields (e.g., 2D polarization) that present periodic textures preserving the sign of the Skyrme density, unavoidably exhibit zeros. We implement these textures in the polarization state of a laser beam.
\end{abstract}

\maketitle


    \textit{Introduction}.\,---
    We refer to a 2D skyrmion as a distribution (or texture) over a plane of a vector field that completely spans 
    a spherical parameter space, while maintaining the sign of the Skyrme density $\rho_\mathrm{S}$ (the Jacobian between the 
    sphere and the planar 
    region occupied by the skyrmion). 
    By definition, the Skyrme number, $N_\mathrm{S}$, given by the integral of $\rho_\mathrm{S}$
    over the region occupied by the skyrmion, is an integer. 
    Merons are distributions that span 
    one hemisphere, north or south, their Skyrme number then being $\pm1/2$ when the hemisphere is spanned once.
    Skyrmions and other 2D textures have been observed in a plethora of physical systems, such as magnetic materials \cite{magnetic_skyrmions_review, beyond_skyrmions, meron_lattices_magnetic}, sound waves \cite{acoustic_skyrmions, Muelas_soundwaves}, superfluids \cite{meron_lattice_superfluid,meron_lattice_superfluid_book} and optical fields \cite{review_optical_skyrmions_Shen}. In nonparaxial optical fields, 
    isolated skyrmions whose vector parameter is the normalized photonic spin angular momentum \cite{skyrmion_spin_evanescent,skyrmions_Rodrigo_Pisanty} have been found. 
    Periodic textures, such as skyrmion lattices for the electric field's instantaneous orientation in evanescent waves \cite{first_optical_skyrmions} and spin optical meron lattices \cite{optical__merons_PRL,optical__merons_Zhang,optical_plasmonic_merons, spin_merons_polygons,Airy_beams_merons} have also been observed.
 
    In paraxial optics, 2D skyrmionic textures 
    in the polarization distribution over the transverse plane of monochromatic beams are often referred to as \textit{Stokes textures} \cite{review_optical_skyrmions_Shen, Shen_bimerons, skyrmions_lens}, since they span all normalized values of the Stokes vector, that is, the Poincaré sphere. Full Poincaré beams \cite{full_Poincare_beams, Poincare_skyrmions, paraxial_skyrmionic_beams}, which display a spatially variant polarization pattern in a transverse section that is the stereographic projection of the Poincaré sphere, can be regarded as isolated skyrmions. 
    
    
    Regarding periodic textures, optical polarization lattices of tiles that separately map each Poincaré hemisphere 
    can be recognized as \textit{Stokes meron lattices}. For example, the superposition of three plane waves with appropriate polarizations and wavevectors results in a triangular lattice \cite{lemon_fields}. Note that
    in this lattice $\rho_\mathrm{S}$ oscillates in sign, as does that in nonparaxial spin optical meron lattices \cite{optical__merons_PRL,optical__merons_Zhang,optical_plasmonic_merons, spin_merons_polygons}. This oscillation may lead to cancellation when integrating $\rho_\mathrm{S}$, yielding $N_\mathrm{S}\approx0$ in some regions. In contrast, a few periodic textures turn out to avoid such cancellation, since they preserve $\mathrm{sgn}(\rho_\mathrm{S})$ over all space, such as optical nonparaxial skyrmion lattices \cite{first_optical_skyrmions}, meron lattices in superfluid \textsuperscript{3}He-A \cite{meron_lattice_superfluid}, or 
    textures in the velocity field of sound waves \cite{Muelas_soundwaves}. We recently proposed propagation-invariant optical polarization lattices that were designed to present this property \cite{prop_invariant_meron_lattices}.

\begin{figure*}
    \centering
    \def\svgwidth{1\textwidth}
    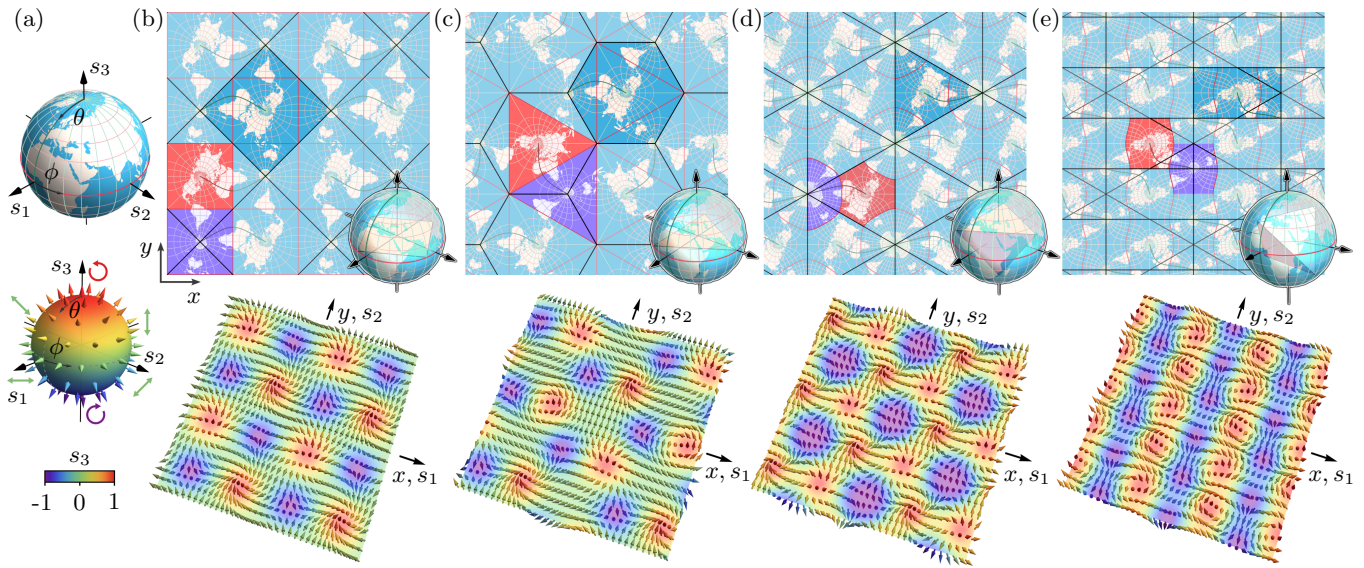\caption{\label{fig:cartography_skyrmions} (a) The Earth and a spanned spherical parameter space (e.g. the Poincaré sphere).
    (b-e) Periodic skyrmionic textures resulting from conformal cartographic mappings providing regular tessellations: (b) Peirce's quincuncial projection, (c) Adams' world in a hexagon, (d) Lee's world in a tetrahedron,
    and (e) Wray's variation of Lee's projection. For all, the entire sphere is mapped onto regular polygons (outlined in black), one of them highlighted in darker blue in each case.  Rotated versions of these tessellate the plane.
    A northern and a southern hemispheres in each case are 
    highlighted in red and purple, respectively.
    }
    \end{figure*}

    The first goal of this work is to study skyrmionic textures that preserve the sign of the Skyrme density. Such textures would result naturally from the accumulation of skyrmions with equal (nonzero) Skyrme number. We show that the subset of these textures for which the mapping from the sphere to the plane is conformal minimize a geometric measure of the energy of the mapping. These textures are then stable when the interaction between skyrmions follows the minimization of this energy.
    As it turns out, the  periodic conformal maps that describe these textures were 
%
    %
    developed in nineteenth- and twentieth-century cartography, 
    where the sphere 
    is mapped 
    onto specific regular polygons. 
    Physical space 
    is then tessellated with these polygons without reflections, only translations and rotations. 
    The spherical parameter space 
    is then mapped as if it were the Earth (Fig.~\ref{fig:cartography_skyrmions}(a)). 
    
    Our second goal is to consider the implementation of these conformal periodic skyrmion lattices 
    as Stokes textures in paraxial monochromatic fields, for which 
    the mapped spherical space is the Poincaré (or Bloch) sphere. 
    We propose 
    a prescription that generates well-behaved fields. 
    We also find that the implementation of periodic skyrmionic structures with uniform ${\rm sgn}(\rho_{\rm S})$ necessarily implies that the field must present zeros, which make the pattern sensitive to perturbations.
    
\textit{Energy minimization}.\,---
The Skyrme number, $N_\textrm{S}$, quantifies how many times the unit vector $\mathbf{s}=(\sin\theta\cos\phi,\sin\theta\sin\phi,\cos\theta)$ wraps around the sphere, its sign giving the sense of wrapping. Within a given region $\sigma$, 
$N_\textrm{S} = (4\pi)^{-1} \iint_\sigma \rho_\textrm{S} (x,y) \; \mathrm{d}x \, \mathrm{d}y$, where \(\rho_\textrm{S} (x,y) = \mathbf{s}(x,y) \cdot \left[\partial_x \mathbf{s}(x,y) \times \partial_y \mathbf{s}(x,y) \right]\) 
is the Jacobian of the transformation. Let us define an energy density of the mapping as a measure of the local stretching from the sphere onto the flat space, namely the Euclidean norm of the $2\times3$ Jacobian matrix: $\epsilon=||\nabla{\bf s}||^2/2$. 
It is shown in the \hyperref[sec:variational_calc]{Supplemental Information, Sec.~1}, that, for a periodic texture with constant ${\rm sgn}(\rho_{\rm S})$, the total energy over a unit cell has as minimum value $4\pi|N_{\rm S}|$, and this minimum energy is achieved when the map is conformal, in which a case $\epsilon=|\rho_{\rm S}|$. Therefore, in physical situations where the dynamics follow minimization of the energy as defined earlier,  the accumulation of large numbers of skyrmions whose Skyrme density has equal sign naturally results in periodic conformal maps.

\textit{Cartographic mappings}.\,---
We now present the conformal cartographic transformations and their implementation as skyrmionic textures (Fig.~\ref{fig:cartography_skyrmions}). 
 Only three tiles provide regular tessellations \cite{Fathauer_tessellations}: 
squares, equilateral triangles and regular hexagons. Here, we explore conformal maps that yield regular tessellations where either the entire sphere or a hemisphere is mapped onto each 
regular polygon. The only case left out is that where a hemisphere is mapped onto a hexagon, since it is impossible to tile the plane without edge conflicts using only rotated copies of hexagonal merons.

We begin with Peirce's {\it quincuncial} projection \cite{Peirce}, which maps the whole sphere, and also each hemisphere, onto a square. The vertices of each square tile mapping a hemisphere are equally spaced along the sphere's equator 
(Fig.~\ref{fig:cartography_skyrmions}(b)). The map is conformal except at these four points. The resulting meron texture is topologically equivalent and morphologically similar to a texture resulting from energy minimization in 
superfluid \textsuperscript{3}He-A \cite{meron_lattice_superfluid}.

We consider next Adams' {\it world in a hexagon} projection \cite{Adams_hexagon}, which maps the sphere onto a regular hexagon, while mapping each hemisphere onto an equilateral triangle. This map exhibits singular points at the vertices of each triangular tile that spans a hemisphere, which are again equally spaced along the equator (Fig.~\ref{fig:cartography_skyrmions}(c)). We recently generated a Stokes texture corresponding to the superposition of a few plane waves that is topologically equivalent to this map, albeit not conformal \cite{prop_invariant_meron_lattices}.

Finally, we explore a 
map of 
the entire sphere 
onto an equilateral triangle, known as Lee's {\it world in a tetrahedron} projection \cite{Lee_tetrahedron}. Here, the sphere is divided into four equal spherical equilateral triangles, each being then projected onto the corresponding face of a regular tetrahedron. Unfolding the tetrahedron yields an equilateral triangle that tessellates the plane with rotated copies of itself. Note that one of the tetrahedron's vertices is chosen to be at the south pole (Fig.~\ref{fig:cartography_skyrmions}(d)). 
This texture is topologically equivalent and morphologically similar to one generated recently in the instantaneous velocity field of sound waves \cite{Muelas_soundwaves}. 
Wray proposed a modified version of Lee's mapping \cite{Lee_Dixon} 
where the two poles are positioned at the midpoints of two opposing edges of the tetrahedron (Fig.~\ref{fig:cartography_skyrmions}(e)) . Unlike Lee's original projection, in Wray's transformation the equator does not form closed loops but meandering lines. 
Upon unfolding the tetrahedron, the resulting triangle can be rearranged as a rectangle (with aspect ratio $\sqrt3:1$), yielding a tessellation of rectangles. In both versions of Lee's projection (Figs.~\ref{fig:cartography_skyrmions}(d,e)), the singular points are placed at the vertices of the tetrahedron.


These maps can be expressed as a sequence of 
two conformal transformations: a planar map defined by an analytic function $w(z)$, where $z=x+\im y$ ($x$ and $y$ being the Cartesian coordinates in physical space), and a stereographic projection from the north pole. These operations result in a map from the plane to
the sphere's azimuthal and polar coordinates as $\phi=\mathrm{arg}\left[w(z)\right]$ and $\theta=2\arctan |w(z)|$, respectively. The 3D Cartesian coordinates in the sphere's ambient space are $s_1=\sin\theta \cos\phi$, $s_2=\sin\theta \sin\phi$ and $s_3=\cos\theta$.

For Peirce's and Adams' projections, $w(z)$ is the inverse function of the Schwarz-Christoffel transformation \cite{Schwarz} that maps the unit disk onto 
a square 
\cite{Lee_Jacobi} and an equilateral triangle 
\cite{Lee_Dixon}, respectively. The expressions for $w(z)$ for each of these projections (Peirce \cite{Lee_Jacobi}, Adams, Lee, and Wray \cite{Lee_Dixon}) are:
\begin{subequations}
\label{eq:maps}
\begin{align}
    w_\mathrm{P} (z)&=\frac{e^{\im (\phi_0 + \alpha)}}{\sqrt{2}} \mathrm{sd}\left(\tilde{z}_1, \frac{1}{\sqrt2} \right), \label{eq:Peirce}
    \\
    w_\mathrm{A} (z)&=e^{\im (\phi_0 + \alpha)} \mathrm{sm}(\tilde{z}_2), \label{eq:Adams}
    \\
    w_\mathrm{L} (z)&=2^{1/6} e^{\im (\phi_0 + \alpha)}
    \mathrm{sm}(\tilde{z}_2) \mathrm{cm}(\tilde{z}_2),\label{eq:Lee}
     \\
    w_\mathrm{W} (z)&=\frac{e^{\im (\phi_0 + \alpha)}}{2} \mathrm{sd}\left(\tilde{z}_3, \frac{\sqrt3-1}{2\sqrt2} \right),\label{eq:Wray}
 \end{align}
 \end{subequations}
where sd is the ratio between the Jacobi elliptic functions sn and dn, while sm and cm correspond to the Dixon elliptic functions, which can be expressed in terms of the Weierstrass elliptic function denoted as $\wp(z)$ and its derivative $\wp'(z)$: $\mathrm{sm}(z)=6 \wp(z)/(1-3 \wp'(z))$ and $\mathrm{cm}(z)=(3\wp'(z)+1)/(3\wp'(z)-1)$ \cite{elliptic_functions_Conrad}. The three complex numbers $\tilde{z}_i$ are defined as $\tilde{z}_{1,2,3}=\gamma_{1,2,3} \; e^{-\im \alpha}(z-z_0)/d$, with $\gamma_1=\Gamma^2(1/4)/(2\sqrt{2 \pi})$, $\gamma_\mathrm{2}=\Gamma^3(1/3)/(2\pi)$, and $\gamma_\mathrm{3}=3^{1/4} \Gamma^3(1/3)/(2^{1/3}\pi)$, where $\Gamma$ denotes the gamma function. 
The angle $\alpha$ rotates the polygon onto which the sphere is mapped, while $\phi_0$ sets the initial value from which $\phi$ is swept within this polygon. Note that $\phi_0$ influences the type of meron that appears in the textures (e.g., Neel- or Bloch-type \cite{review_optical_skyrmions_Shen}). The real and imaginary parts of the complex number $z_0$ are the $x$ and $y$ coordinates defining the center of the map. Finally, $d$ sets the length of the edge of either the square or the equilateral triangle spanning a hemisphere in the case of Peirce's and Adams' mappings, while for Lee's and Wray's mappings it corresponds to the length of the edge of the triangle spanning the sphere.
Note that, due to the periodicity of the Jacobi and Dixon elliptic functions, these mappings work over the whole plane and not only within a central tile. The periodicity of the tiling is then natural.
The inverse transformations to each of the mappings in Eqs.~\eqref{eq:maps} are given in the \hyperref[sec:inverse_mappings]{Supplemental Information, Sec.~2}.

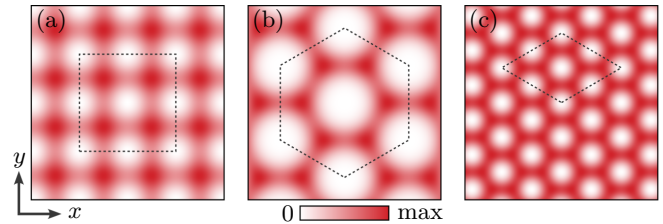
\begin{figure}[h]
    \centering
    \def\svgwidth{0.48\textwidth}
\begingroup%
  \makeatletter%
  \providecommand\color[2][]{%
    \errmessage{(Inkscape) Color is used for the text in Inkscape, but the package 'color.sty' is not loaded}%
    \renewcommand\color[2][]{}%
  }%
  \providecommand\transparent[1]{%
    \errmessage{(Inkscape) Transparency is used (non-zero) for the text in Inkscape, but the package 'transparent.sty' is not loaded}%
    \renewcommand\transparent[1]{}%
  }%
  \providecommand\rotatebox[2]{#2}%
  \newcommand*\fsize{\dimexpr\f@size pt\relax}%
  \newcommand*\lineheight[1]{\fontsize{\fsize}{#1\fsize}\selectfont}%
  \ifx\svgwidth\undefined%
    \setlength{\unitlength}{657.63779528bp}%
    \ifx\svgscale\undefined%
      \relax%
    \else%
      \setlength{\unitlength}{\unitlength * \real{\svgscale}}%
    \fi%
  \else%
    \setlength{\unitlength}{\svgwidth}%
  \fi%
  \global\let\svgwidth\undefined%
  \global\let\svgscale\undefined%
  \makeatother%
  \begin{picture}(1,0.33313363)%
    \lineheight{1}%
    \setlength\tabcolsep{0pt}%
    \put(0,0){\includegraphics[width=\unitlength,page=1]{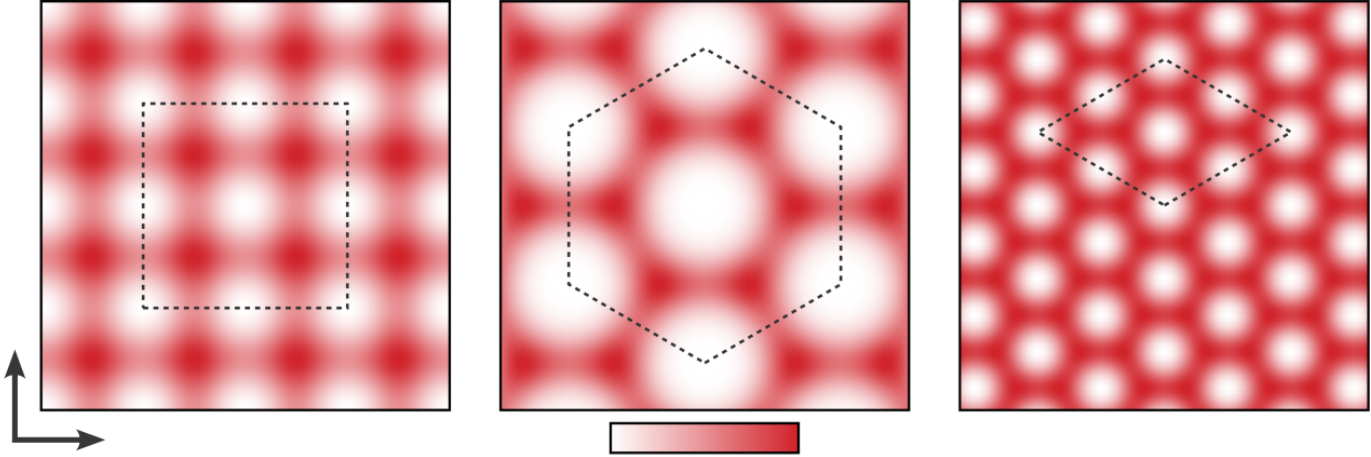}}%
    \put(0.03708873,0.29979065){\color[rgb]{0,0,0}\makebox(0,0)[lt]{\lineheight{1.25}\smash{\begin{tabular}[t]{l}(a)\end{tabular}}}}%
    \put(0.3715502,0.29979065){\color[rgb]{0,0,0}\makebox(0,0)[lt]{\lineheight{1.25}\smash{\begin{tabular}[t]{l}(b)\end{tabular}}}}%
    \put(0.70641891,0.29979065){\color[rgb]{0,0,0}\makebox(0,0)[lt]{\lineheight{1.25}\smash{\begin{tabular}[t]{l}(c)\end{tabular}}}}%
    \put(0.5960389,0.00322703){\color[rgb]{0,0,0}\makebox(0,0)[lt]{\lineheight{1.25}\smash{\begin{tabular}[t]{l}\small{max}\end{tabular}}}}%
    \put(0.00233835,0.09069117){\color[rgb]{0,0,0}\makebox(0,0)[lt]{\lineheight{1.25}\smash{\begin{tabular}[t]{l}$y$\end{tabular}}}}%
    \put(0.0872371,0.00520308){\color[rgb]{0,0,0}\makebox(0,0)[lt]{\lineheight{1.25}\smash{\begin{tabular}[t]{l}$x$\end{tabular}}}}%
    \put(0.41718527,0.00041603){\color[rgb]{0,0,0}\makebox(0,0)[lt]{\lineheight{1.25}\smash{\begin{tabular}[t]{l}\small{0}\end{tabular}}}}%
  \end{picture}%
\endgroup%
\caption{\label{fig:Skyrme_density} Skyrme density
    for (a) Peirce, (b) Adams, and (c) Lee/Wray maps within the area displayed in Fig.~\ref{fig:cartography_skyrmions}. The unit cell of each texture is outlined by dashed lines.
    }
    \end{figure}

The Skyrme number 
can be expressed as $\rho_\mathrm{S} = \rho_\mathrm{SP} \rho_w$, where $\rho_\mathrm{SP} =4(1+|w(z)|^2)^{-2}$ is the Jacobian of the stereographic projection, and $ \rho_w =\left[\partial_x \mathrm{Re}(w(z))\right]^2+\left[\partial_y \mathrm{Re}(w(z)) \right]^2$ is that for the transformation $w(z)$ (simplified by using the Cauchy-Riemann conditions). These equations imply that $\rho_\mathrm{S}$ is always positive, as shown in Fig.~\ref{fig:Skyrme_density}. Within the context of cartography, this means that the shapes of the continents are never reversed. 
A different convention for the stereographic projection could be used that reverses its sign, leading to textures with $\mathrm{sgn}(\rho_\mathrm{S})<0$.

    Note that $N_\textrm{S}=1$ within the polygons that map the entire sphere for all projections in Fig.~\ref{fig:cartography_skyrmions}. Peirce's and Adams' projections lead to lattices of square and triangular merons, respectively, in which any two neighboring merons map different hemispheres. 
    Within each meron, $N_\textrm{S}=1/2$.
    Nevertheless, in the first version of Lee's projection (Fig.~\ref{fig:cartography_skyrmions}(d)), there are merons that span twice the southern hemisphere, leading to $N_\textrm{S}=1$. In fact, both in Lee's and Wray's versions of the tetrahedron projection, different regions covering the same hemisphere are in direct contact and not separated by a line corresponding to the equator, so these textures do not classify as meron lattices according to some definitions \cite{magnetic_skyrmions_review}.
      

\textit{Periodic skyrmionic textures in monochromatic paraxial optical fields}.\,--- We now explore the implementation of these skyrmionic textures in the spatial distribution of polarization for a monochromatic optical beam of frequency $\omega$. The transverse electric field at any point $(x,y)$ and time $t$ is given by $\boldsymbol{\mathcal{E}}=\mathrm{Re}\left[\mathbf{E} \; \mathrm{exp}(-\im\omega t)\right]$, where $\mathbf{E}$ is the complex field, which can be parametrized as
$\mathbf{E}=E_0\left[\cos(\theta/2)\mathbf{l} + e^{\im\phi}\sin(\theta/2)\mathbf{r}\right]=E_\mathbf{l} \mathbf{l} + E_\mathbf{r} \mathbf{r}$,
with $E_0$ being a complex function of $x$ and $y$ that provides an overall amplitude and phase and that does not affect the polarization (defined by the expression in brackets). Here, the unit vectors $\mathbf{l}, \mathbf{r}=(\mathbf{x}\pm \im \mathbf{y})/\sqrt2$ represent left- and right-circular polarization, where $\mathbf{x}$ and $\mathbf{y}$ are unit vectors along the $x$ and $y$ axes, respectively. The circular polarization components $E_{\mathbf{l}, \mathbf{r}}$ are complex functions of $x$ and $y$, whose relative phase is
$\phi=\phi_\mathbf{r}-\phi_\mathbf{l}$ (with $\phi_{\mathbf{r},\mathbf{l}}=\mathrm{arg}[E_{\mathbf{r},\mathbf{l}}]$). 
At each point, $\boldsymbol{\mathcal{E}}$ traces over each optical cycle an ellipse within the transverse plane, whose ellipticity and handedness is determined by $\theta$, while $\phi$ gives twice the orientation angle of its major axis with respect to $x$. Each possible polarization ellipse then corresponds to a point on the Poincaré sphere (Fig.~\ref{fig:cartography_skyrmions}(a)) described by a normalized Stokes vector $\mathbf{s}=(\sin\theta\cos\phi,\sin\theta\sin\phi,\cos\theta)$. Spanning the Poincaré sphere then means achieving every paraxial polarization. The northern and southern hemispheres correspond to left- and right-handed ellipses, respectively, with the two poles corresponding to the two circular polarizations, while linearly polarized ellipses are positioned along the equator.

\begin{figure}[h]
\centering
\def\svgwidth{0.48\textwidth}
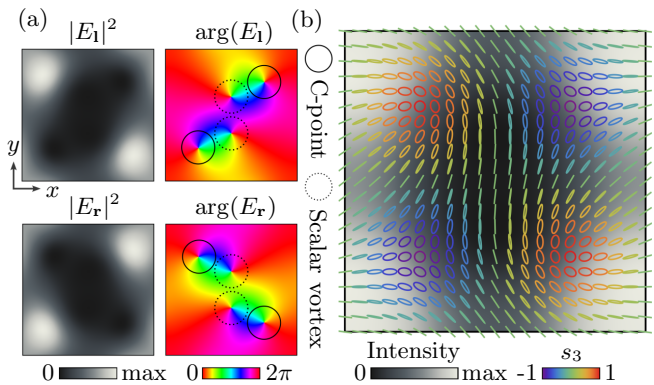\caption{\label{fig:Peirce_field_theory}
Unit cell of the 
beam implementing in its polarization state Peirce's texture according to Eqs.~\eqref{eq:maps} with $\alpha=\pi/4$, $\phi_0=7\pi/6$, and $z_0=(-1+\im)/2$. (a) Theoretical intensity and phase distributions of the circular polarization components $E_{\mathbf{l},\mathbf{r}}$. (b) Polarization and intensity distributions. Phase vortices result in either C-points or scalar vortices.
    }
    \end{figure}

We now derive equations for optical fields that implement these textures, independently of the mapping function $w(z)$. Since $w(z)=\tan(\theta/2)e^{\im\phi}$,
$\mathbf{E}=E_\mathbf{l}(z)[\mathbf{l}+w(z)\mathbf{r}]$,
so any reasonable choice for $E_\mathbf{l}(z)$ leads to the desired polarization texture as long as $E_\mathbf{r}(z)=E_\mathbf{l}(z)w(z)$. However, it is convenient to choose a functional form that avoids singularities in the field, and that treats the two hemispheres similarly. In our convention, the hemisphere with left(right)-handed ellipses corresponds to $|w(z)|<(>)\,1$, with the pole representing left(right)-circular polarization corresponding to $w(z)=0\,(\infty)$. It is then desirable to find an expression for $E_\mathbf{l}(z)$ such that $E_\mathbf{r}(z)=E_\mathbf{l}(z)w(z)$ takes a similar functional form except for the replacement $w\to w^{-1}$. One such solution is
%
%
\begin{eqnarray}
\label{components}
E_{\mathbf{l},\mathbf{r}} (z)&=&{\cal E}\sqrt{-w^{\pm1}_0}\,\frac{[w^{\pm1}(z)-w_0^{\pm1}]^*}{|w^{\pm1}(z)|^2+|w_0^{\pm1}|^2}, 
%
\end{eqnarray}
%
where ${\cal E}$ is a constant amplitude factor and $w_0$ is a dimensionless constant. Note that both polarization components vanish at points where $w(z)=w_0$, meaning that this construction forces zeros in the field at points where the polarization is such that $\tan(\theta/2)e^{\im\phi}=w_0$. As we discuss later, this 
feature has 
a topological origin. 
In what follows we choose $w_0=-1$ to simplify the expressions, so that the zeros coincide with points of vertical linear polarization.
We now focus on Peirce's texture as a representative example, for which 
Fig.~\ref{fig:Peirce_field_theory} shows the intensity and phase of $E_{\mathbf{l},\mathbf{r}}$ (a), as well as the polarization and intensity distributions (b). 
The figures for the other maps are shown
in the \hyperref[sec:additional_theory]{Supplemental Material, Sec.~3}, which also includes plots of the fields' discrete Fourier spectra for the four maps.

These textures correspond to optical polarization lattices \cite{Freund_lattices} exhibiting polarization singularities such as L-lines (lines of linear polarization) and C-points (points of circular polarization) \cite{Dennis_singularities}. 
C-points emerge in regions where a vortex in $\phi_{\mathbf{l},\mathbf{r}}$ with topological charge $m_{\mathbf{l},\mathbf{r}}$ is superimposed onto a region with no vortex in $\phi_{\mathbf{r},\mathbf{l}}$, rendering a vortex in $\phi$ with topological charge $m_\phi=\mp m_{\mathbf{l},\mathbf{r}}$.
The polarization ellipse undergoes a rotation of $m_\phi \pi$ radians around the singularity in $\phi$.
For $m_\phi>0$, the ellipse performs a counterclockwise rotation, while for $m_\phi<0$ it rotates clockwise. In the proximity of C-points with $m_\phi = \pm1$, the ellipses exhibit distinct singular patterns known as lemons ($m_\phi = 1$) and stars ($m_\phi = -1$) \cite{Dennis_singularities}.

The polarization distribution in Fig.~\ref{fig:Peirce_field_theory}(b) shows an important feature: For a texture where $\rho_\mathrm{S} > 0$ everywhere, all left(right)-handed C-points originate from a $\phi$ vortex with $m_\phi >(<)\,0$, and then all C-points with $m_\phi = \pm1$ manifest as left-handed lemons and right-handed stars (see discussion in \hyperref[sec:topological_properties]{Supplemental Material, Sec.~4}). (For a texture where $\rho_\mathrm{S} < 0$ everywhere the situation would reverse.)
Together with a well-established result from index theory, this result
implies that paraxial fields displaying periodic skyrmionic Stokes textures that maintain $\mathrm{sgn}(\rho_\mathrm{S})$
must inevitably present zeros. This is because, within a unit cell of a periodic complex scalar field, the topological charges of its vortices must add up to zero. However, as just discussed, the vortices for each circular component that lead to C-points all have the same charge. Each component must then include extra vortices with the opposite charge, and the only way for these not to produce extra C-points is if their locations coincide for the two circular components, hence giving rise to field zeros. The field construction in Eqs.~\eqref{components} naturally introduces these zeros, and places them at points with a specific polarization. 

It is important to emphasize that this 
result extends beyond textures implemented in paraxial polarization; it 
applies to any texture in a 2D spinor field achieved through the factorization of a global phase and amplitude function in a 2D complex vector field. Consequently, in such fields a periodic texture that preserves $\mathrm{sgn}(\rho_\mathrm{S})$ inevitably presents zeros.

\begin{figure}
\centering
\def\svgwidth{0.48\textwidth}
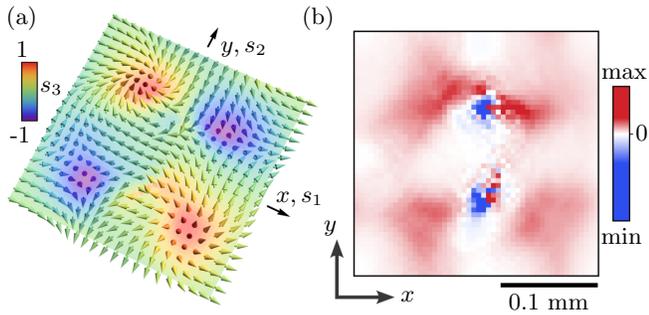\caption{\label{fig:Peirce_experiment}
Unit cell of the measured (a) Stokes vector distribution and (b) Skyrme density for Peirce's field.}
\end{figure}

We generated the textures experimentally by implementing the circular polarization distributions in Eq.~(\ref{components}) in a laser beam using a spatial light modulator \cite{vector_setup,method_modulation} (see \hyperref[sec:experiment]{Supplemental Information, Sec.~5}).
Figure~\ref{fig:Peirce_experiment} shows, for Peirce's projection, the measured Stokes vector distribution (a) and $\rho_\mathrm{S}$ (b). Similar results for the other three projections are given in the \hyperref[sec:additional_experiments]{Supplemental Material, Sec.~6}. For all cases, small regions of highly negative $\rho_\mathrm{S}$ emerge near the points where Eq.~\eqref{components} predicts the field's zeros
(corresponding to vertical polarization). This is because these zeros are unstable: any small relative misalignment of the polarization components causes the intensity not to vanish exactly and the polarization to vary rapidly, covering within a small region the entire sphere in the opposite sense.
Possible experimental sources of zero misalignment are defocus or the clipping of diffraction orders by the aperture. Excluding some orders carrying low intensities has little effect on the overall pattern but leads to rapid polarization variation in low-intensity regions (see \hyperref[sec:instability]{Supplemental Information, Sec.~7}), similar to that in the measured data.

\textit{Conclusions}.\,--- The skyrmionic lattices presented here represent different morphologies and topologies that preserve $\mathrm{sgn}(\rho_\mathrm{S})$. The fact that they are conformal implies that they minimize the energy inherent to the mapping.
Note that rotated versions of each of these cartographic maps can also be considered, like the two forms of Lee's projection. Similarly, 
some of these maps accept deformations while still tessellating the plane but not with regular polygons, and perhaps not conformally. It is possible, for example, to deform continuously Wray's map by transporting the singular points along meridians to the equator, arriving at a rotated version of Peirce's map. This means that Lee's, Wray's and Peirce's maps are topologically equivalent. Note, however, that changing the spacing of the singular points in Peirce's map until two of them merge would not lead to Adams' map (and would not be consistent with tesselation of the plane), meaning that this latter map (with three singular points) is topologically different from the others (with four singular points).

Other conformal projections that tessellate the plane can also be explored \cite{Lee_Dixon, Lee_Jacobi}, such as maps of the sphere onto a triangle or a square using Lagrange's projection of the sphere onto a disk \cite{Lee_circle} instead of the stereographic projection. An interesting case is Adams' {\it world in a square} projection \cite{Adams_square}, which renders merons spanning a hemisphere four times for both hemispheres. Another texture with merons spanning a hemisphere six times for one of the hemispheres results from Cox' {\it world in a triangle} projection \cite{Cox}.

For polarization textures (or any spinor field resulting from a 2D complex vector field), achieving uniform $\mathrm{sgn}(\rho_\mathrm{S})$ requires zeros in the vector field, making the texture unstable to perturbations, although instability under propagation can be reduced in some cases \cite{prop_invariant_meron_lattices}. Exploring the inevitable occurrence of zeros in spinor fields beyond optics, such as Bose-Einstein condensates \cite{hansen2016singular}, opens up an area for further investigation.

Finally, it would be interesting to study if periodic textures with uniform $\mathrm{sgn}(\rho_\mathrm{S})$,
conformal or not, exist for higher dimensionalities. Recently, optical implementations of skyrmionic structures spanning 3-spheres \cite{sugic2021particle} and 4-spheres \cite{F3DP_fields} were proposed, the latter corresponding to a space-time periodic texture spanning all states of monochromatic nonparaxial polarization, but with nonuniform
$\mathrm{sgn}(\rho_\mathrm{S})$.

\begin{acknowledgments}
This research received funding from the Agence Nationale de Recherche (ANR) through the project 3DPol, ANR-21-CE24-0014-01. D. M. acknowledges Ministerio de Universidades, Spain, Universidad Miguel Hernández and the European Union (Next generation EU fund) for a Margarita Salas grant
from the program Ayudas para la Recualificación del Sistema Universitario Español, and funding from Ministerio de Ciencia e Innovación, Spain, PID2021-126509OB-C22.
\end{acknowledgments}
 
\bibliography{apssamp}

\providecommand{\noopsort}[1]{}\providecommand{\singleletter}[1]{#1}%
\begin{thebibliography}{41}%
\makeatletter
\providecommand \@ifxundefined [1]{%
 \@ifx{#1\undefined}
}%
\providecommand \@ifnum [1]{%
 \ifnum #1\expandafter \@firstoftwo
 \else \expandafter \@secondoftwo
 \fi
}%
\providecommand \@ifx [1]{%
 \ifx #1\expandafter \@firstoftwo
 \else \expandafter \@secondoftwo
 \fi
}%
\providecommand \natexlab [1]{#1}%
\providecommand \enquote  [1]{``#1''}%
\providecommand \bibnamefont  [1]{#1}%
\providecommand \bibfnamefont [1]{#1}%
\providecommand \citenamefont [1]{#1}%
\providecommand \href@noop [0]{\@secondoftwo}%
\providecommand \href [0]{\begingroup \@sanitize@url \@href}%
\providecommand \@href[1]{\@@startlink{#1}\@@href}%
\providecommand \@@href[1]{\endgroup#1\@@endlink}%
\providecommand \@sanitize@url [0]{\catcode `\\12\catcode `\$12\catcode `\&12\catcode `\#12\catcode `\^12\catcode `\_12\catcode `\%12\relax}%
\providecommand \@@startlink[1]{}%
\providecommand \@@endlink[0]{}%
\providecommand \url  [0]{\begingroup\@sanitize@url \@url }%
\providecommand \@url [1]{\endgroup\@href {#1}{\urlprefix }}%
\providecommand \urlprefix  [0]{URL }%
\providecommand \Eprint [0]{\href }%
\providecommand \doibase [0]{https://doi.org/}%
\providecommand \selectlanguage [0]{\@gobble}%
\providecommand \bibinfo  [0]{\@secondoftwo}%
\providecommand \bibfield  [0]{\@secondoftwo}%
\providecommand \translation [1]{[#1]}%
\providecommand \BibitemOpen [0]{}%
\providecommand \bibitemStop [0]{}%
\providecommand \bibitemNoStop [0]{.\EOS\space}%
\providecommand \EOS [0]{\spacefactor3000\relax}%
\providecommand \BibitemShut  [1]{\csname bibitem#1\endcsname}%
\let\auto@bib@innerbib\@empty
\bibitem [{\citenamefont {Nagaosa}\ and\ \citenamefont {Tokura}(2013)}]{magnetic_skyrmions_review}%
  \BibitemOpen
  \bibfield  {author} {\bibinfo {author} {\bibfnamefont {N.}~\bibnamefont {Nagaosa}}\ and\ \bibinfo {author} {\bibfnamefont {Y.}~\bibnamefont {Tokura}},\ }\bibfield  {title} {\bibinfo {title} {Topological properties and dynamics of magnetic skyrmions},\ }\href@noop {} {\bibfield  {journal} {\bibinfo  {journal} {Nature Nanotechnology}\ }\textbf {\bibinfo {volume} {8}},\ \bibinfo {pages} {899} (\bibinfo {year} {2013})}\BibitemShut {NoStop}%
\bibitem [{\citenamefont {G{\"o}bel}\ \emph {et~al.}(2021)\citenamefont {G{\"o}bel}, \citenamefont {Mertig},\ and\ \citenamefont {Tretiakov}}]{beyond_skyrmions}%
  \BibitemOpen
  \bibfield  {author} {\bibinfo {author} {\bibfnamefont {B.}~\bibnamefont {G{\"o}bel}}, \bibinfo {author} {\bibfnamefont {I.}~\bibnamefont {Mertig}},\ and\ \bibinfo {author} {\bibfnamefont {O.~A.}\ \bibnamefont {Tretiakov}},\ }\bibfield  {title} {\bibinfo {title} {Beyond skyrmions: Review and perspectives of alternative magnetic quasiparticles},\ }\href@noop {} {\bibfield  {journal} {\bibinfo  {journal} {Physics Reports}\ }\textbf {\bibinfo {volume} {895}},\ \bibinfo {pages} {1} (\bibinfo {year} {2021})}\BibitemShut {NoStop}%
\bibitem [{\citenamefont {Yu}\ \emph {et~al.}(2018)\citenamefont {Yu}, \citenamefont {Koshibae}, \citenamefont {Tokunaga}, \citenamefont {Shibata}, \citenamefont {Taguchi}, \citenamefont {Nagaosa},\ and\ \citenamefont {Tokura}}]{meron_lattices_magnetic}%
  \BibitemOpen
  \bibfield  {author} {\bibinfo {author} {\bibfnamefont {X.}~\bibnamefont {Yu}}, \bibinfo {author} {\bibfnamefont {W.}~\bibnamefont {Koshibae}}, \bibinfo {author} {\bibfnamefont {Y.}~\bibnamefont {Tokunaga}}, \bibinfo {author} {\bibfnamefont {K.}~\bibnamefont {Shibata}}, \bibinfo {author} {\bibfnamefont {Y.}~\bibnamefont {Taguchi}}, \bibinfo {author} {\bibfnamefont {N.}~\bibnamefont {Nagaosa}},\ and\ \bibinfo {author} {\bibfnamefont {Y.}~\bibnamefont {Tokura}},\ }\bibfield  {title} {\bibinfo {title} {Transformation between meron and skyrmion topological spin textures in a chiral magnet},\ }\href@noop {} {\bibfield  {journal} {\bibinfo  {journal} {Nature}\ }\textbf {\bibinfo {volume} {564}},\ \bibinfo {pages} {95} (\bibinfo {year} {2018})}\BibitemShut {NoStop}%
\bibitem [{\citenamefont {Ge}\ \emph {et~al.}(2021)\citenamefont {Ge}, \citenamefont {Xu}, \citenamefont {Liu}, \citenamefont {Xu}, \citenamefont {Lin}, \citenamefont {Yu}, \citenamefont {Bao}, \citenamefont {Jiang}, \citenamefont {Lu},\ and\ \citenamefont {Chen}}]{acoustic_skyrmions}%
  \BibitemOpen
  \bibfield  {author} {\bibinfo {author} {\bibfnamefont {H.}~\bibnamefont {Ge}}, \bibinfo {author} {\bibfnamefont {X.-Y.}\ \bibnamefont {Xu}}, \bibinfo {author} {\bibfnamefont {L.}~\bibnamefont {Liu}}, \bibinfo {author} {\bibfnamefont {R.}~\bibnamefont {Xu}}, \bibinfo {author} {\bibfnamefont {Z.-K.}\ \bibnamefont {Lin}}, \bibinfo {author} {\bibfnamefont {S.-Y.}\ \bibnamefont {Yu}}, \bibinfo {author} {\bibfnamefont {M.}~\bibnamefont {Bao}}, \bibinfo {author} {\bibfnamefont {J.-H.}\ \bibnamefont {Jiang}}, \bibinfo {author} {\bibfnamefont {M.-H.}\ \bibnamefont {Lu}},\ and\ \bibinfo {author} {\bibfnamefont {Y.-F.}\ \bibnamefont {Chen}},\ }\bibfield  {title} {\bibinfo {title} {Observation of acoustic skyrmions},\ }\href@noop {} {\bibfield  {journal} {\bibinfo  {journal} {Physical Review Letters}\ }\textbf {\bibinfo {volume} {127}},\ \bibinfo {pages} {144502} (\bibinfo {year} {2021})}\BibitemShut {NoStop}%
\bibitem [{\citenamefont {Muelas-Hurtado}\ \emph {et~al.}(2022)\citenamefont {Muelas-Hurtado}, \citenamefont {Volke-Sep{\'u}lveda}, \citenamefont {Ealo}, \citenamefont {Nori}, \citenamefont {Alonso}, \citenamefont {Bliokh},\ and\ \citenamefont {Brasselet}}]{Muelas_soundwaves}%
  \BibitemOpen
  \bibfield  {author} {\bibinfo {author} {\bibfnamefont {R.~D.}\ \bibnamefont {Muelas-Hurtado}}, \bibinfo {author} {\bibfnamefont {K.}~\bibnamefont {Volke-Sep{\'u}lveda}}, \bibinfo {author} {\bibfnamefont {J.~L.}\ \bibnamefont {Ealo}}, \bibinfo {author} {\bibfnamefont {F.}~\bibnamefont {Nori}}, \bibinfo {author} {\bibfnamefont {M.~A.}\ \bibnamefont {Alonso}}, \bibinfo {author} {\bibfnamefont {K.~Y.}\ \bibnamefont {Bliokh}},\ and\ \bibinfo {author} {\bibfnamefont {E.}~\bibnamefont {Brasselet}},\ }\bibfield  {title} {\bibinfo {title} {Observation of polarization singularities and topological textures in sound waves},\ }\href@noop {} {\bibfield  {journal} {\bibinfo  {journal} {Physical Review Letters}\ }\textbf {\bibinfo {volume} {129}},\ \bibinfo {pages} {204301} (\bibinfo {year} {2022})}\BibitemShut {NoStop}%
\bibitem [{\citenamefont {Fujita}\ \emph {et~al.}(1978)\citenamefont {Fujita}, \citenamefont {Nakahara}, \citenamefont {Ohmi},\ and\ \citenamefont {Tsuneto}}]{meron_lattice_superfluid}%
  \BibitemOpen
  \bibfield  {author} {\bibinfo {author} {\bibfnamefont {T.}~\bibnamefont {Fujita}}, \bibinfo {author} {\bibfnamefont {M.}~\bibnamefont {Nakahara}}, \bibinfo {author} {\bibfnamefont {T.}~\bibnamefont {Ohmi}},\ and\ \bibinfo {author} {\bibfnamefont {T.}~\bibnamefont {Tsuneto}},\ }\bibfield  {title} {\bibinfo {title} {Textures in rotating superfluid 3{H}e-{A}},\ }\href@noop {} {\bibfield  {journal} {\bibinfo  {journal} {Progress of Theoretical Physics}\ }\textbf {\bibinfo {volume} {60}},\ \bibinfo {pages} {671} (\bibinfo {year} {1978})}\BibitemShut {NoStop}%
\bibitem [{\citenamefont {Bunkov}\ and\ \citenamefont {Godfrin}(2000)}]{meron_lattice_superfluid_book}%
  \BibitemOpen
  \bibfield  {author} {\bibinfo {author} {\bibfnamefont {Y.~M.}\ \bibnamefont {Bunkov}}\ and\ \bibinfo {author} {\bibfnamefont {H.}~\bibnamefont {Godfrin}},\ }\href@noop {} {\emph {\bibinfo {title} {Topological defects and the non-equilibrium dynamics of symmetry breaking phase transitions}}},\ Vol.\ \bibinfo {volume} {549}\ (\bibinfo  {publisher} {Springer Science \& Business Media},\ \bibinfo {year} {2000})\ pp.\ \bibinfo {pages} {325--344}\BibitemShut {NoStop}%
\bibitem [{\citenamefont {Shen}\ \emph {et~al.}(2024{\natexlab{a}})\citenamefont {Shen}, \citenamefont {Zhang}, \citenamefont {Shi}, \citenamefont {Du}, \citenamefont {Yuan},\ and\ \citenamefont {Zayats}}]{review_optical_skyrmions_Shen}%
  \BibitemOpen
  \bibfield  {author} {\bibinfo {author} {\bibfnamefont {Y.}~\bibnamefont {Shen}}, \bibinfo {author} {\bibfnamefont {Q.}~\bibnamefont {Zhang}}, \bibinfo {author} {\bibfnamefont {P.}~\bibnamefont {Shi}}, \bibinfo {author} {\bibfnamefont {L.}~\bibnamefont {Du}}, \bibinfo {author} {\bibfnamefont {X.}~\bibnamefont {Yuan}},\ and\ \bibinfo {author} {\bibfnamefont {A.~V.}\ \bibnamefont {Zayats}},\ }\bibfield  {title} {\bibinfo {title} {Optical skyrmions and other topological quasiparticles of light},\ }\href@noop {} {\bibfield  {journal} {\bibinfo  {journal} {Nature Photonics}\ }\textbf {\bibinfo {volume} {18}},\ \bibinfo {pages} {15} (\bibinfo {year} {2024}{\natexlab{a}})}\BibitemShut {NoStop}%
\bibitem [{\citenamefont {Du}\ \emph {et~al.}(2019)\citenamefont {Du}, \citenamefont {Yang}, \citenamefont {Zayats},\ and\ \citenamefont {Yuan}}]{skyrmion_spin_evanescent}%
  \BibitemOpen
  \bibfield  {author} {\bibinfo {author} {\bibfnamefont {L.}~\bibnamefont {Du}}, \bibinfo {author} {\bibfnamefont {A.}~\bibnamefont {Yang}}, \bibinfo {author} {\bibfnamefont {A.~V.}\ \bibnamefont {Zayats}},\ and\ \bibinfo {author} {\bibfnamefont {X.}~\bibnamefont {Yuan}},\ }\bibfield  {title} {\bibinfo {title} {Deep-subwavelength features of photonic skyrmions in a confined electromagnetic field with orbital angular momentum},\ }\href@noop {} {\bibfield  {journal} {\bibinfo  {journal} {Nature Physics}\ }\textbf {\bibinfo {volume} {15}},\ \bibinfo {pages} {650} (\bibinfo {year} {2019})}\BibitemShut {NoStop}%
\bibitem [{\citenamefont {Guti{\'e}rrez-Cuevas}\ and\ \citenamefont {Pisanty}(2021)}]{skyrmions_Rodrigo_Pisanty}%
  \BibitemOpen
  \bibfield  {author} {\bibinfo {author} {\bibfnamefont {R.}~\bibnamefont {Guti{\'e}rrez-Cuevas}}\ and\ \bibinfo {author} {\bibfnamefont {E.}~\bibnamefont {Pisanty}},\ }\bibfield  {title} {\bibinfo {title} {Optical polarization skyrmionic fields in free space},\ }\href@noop {} {\bibfield  {journal} {\bibinfo  {journal} {Journal of Optics}\ }\textbf {\bibinfo {volume} {23}},\ \bibinfo {pages} {024004} (\bibinfo {year} {2021})}\BibitemShut {NoStop}%
\bibitem [{\citenamefont {Tsesses}\ \emph {et~al.}(2018)\citenamefont {Tsesses}, \citenamefont {Ostrovsky}, \citenamefont {Cohen}, \citenamefont {Gjonaj}, \citenamefont {Lindner},\ and\ \citenamefont {Bartal}}]{first_optical_skyrmions}%
  \BibitemOpen
  \bibfield  {author} {\bibinfo {author} {\bibfnamefont {S.}~\bibnamefont {Tsesses}}, \bibinfo {author} {\bibfnamefont {E.}~\bibnamefont {Ostrovsky}}, \bibinfo {author} {\bibfnamefont {K.}~\bibnamefont {Cohen}}, \bibinfo {author} {\bibfnamefont {B.}~\bibnamefont {Gjonaj}}, \bibinfo {author} {\bibfnamefont {N.}~\bibnamefont {Lindner}},\ and\ \bibinfo {author} {\bibfnamefont {G.}~\bibnamefont {Bartal}},\ }\bibfield  {title} {\bibinfo {title} {Optical skyrmion lattice in evanescent electromagnetic fields},\ }\href@noop {} {\bibfield  {journal} {\bibinfo  {journal} {Science}\ }\textbf {\bibinfo {volume} {361}},\ \bibinfo {pages} {993} (\bibinfo {year} {2018})}\BibitemShut {NoStop}%
\bibitem [{\citenamefont {Lei}\ \emph {et~al.}(2021)\citenamefont {Lei}, \citenamefont {Yang}, \citenamefont {Shi}, \citenamefont {Xie}, \citenamefont {Du}, \citenamefont {Zayats},\ and\ \citenamefont {Yuan}}]{optical__merons_PRL}%
  \BibitemOpen
  \bibfield  {author} {\bibinfo {author} {\bibfnamefont {X.}~\bibnamefont {Lei}}, \bibinfo {author} {\bibfnamefont {A.}~\bibnamefont {Yang}}, \bibinfo {author} {\bibfnamefont {P.}~\bibnamefont {Shi}}, \bibinfo {author} {\bibfnamefont {Z.}~\bibnamefont {Xie}}, \bibinfo {author} {\bibfnamefont {L.}~\bibnamefont {Du}}, \bibinfo {author} {\bibfnamefont {A.~V.}\ \bibnamefont {Zayats}},\ and\ \bibinfo {author} {\bibfnamefont {X.}~\bibnamefont {Yuan}},\ }\bibfield  {title} {\bibinfo {title} {Photonic spin lattices: symmetry constraints for skyrmion and meron topologies},\ }\href@noop {} {\bibfield  {journal} {\bibinfo  {journal} {Physical Review Letters}\ }\textbf {\bibinfo {volume} {127}},\ \bibinfo {pages} {237403} (\bibinfo {year} {2021})}\BibitemShut {NoStop}%
\bibitem [{\citenamefont {Zhang}\ \emph {et~al.}(2022)\citenamefont {Zhang}, \citenamefont {Xie}, \citenamefont {Shi}, \citenamefont {Yang}, \citenamefont {He}, \citenamefont {Du},\ and\ \citenamefont {Yuan}}]{optical__merons_Zhang}%
  \BibitemOpen
  \bibfield  {author} {\bibinfo {author} {\bibfnamefont {Q.}~\bibnamefont {Zhang}}, \bibinfo {author} {\bibfnamefont {Z.}~\bibnamefont {Xie}}, \bibinfo {author} {\bibfnamefont {P.}~\bibnamefont {Shi}}, \bibinfo {author} {\bibfnamefont {H.}~\bibnamefont {Yang}}, \bibinfo {author} {\bibfnamefont {H.}~\bibnamefont {He}}, \bibinfo {author} {\bibfnamefont {L.}~\bibnamefont {Du}},\ and\ \bibinfo {author} {\bibfnamefont {X.}~\bibnamefont {Yuan}},\ }\bibfield  {title} {\bibinfo {title} {Optical topological lattices of {B}loch-type skyrmion and meron topologies},\ }\href@noop {} {\bibfield  {journal} {\bibinfo  {journal} {Photonics Research}\ }\textbf {\bibinfo {volume} {10}},\ \bibinfo {pages} {947} (\bibinfo {year} {2022})}\BibitemShut {NoStop}%
\bibitem [{\citenamefont {Ghosh}\ \emph {et~al.}(2021)\citenamefont {Ghosh}, \citenamefont {Yang}, \citenamefont {Dai}, \citenamefont {Zhou}, \citenamefont {Wang}, \citenamefont {Huang},\ and\ \citenamefont {Petek}}]{optical_plasmonic_merons}%
  \BibitemOpen
  \bibfield  {author} {\bibinfo {author} {\bibfnamefont {A.}~\bibnamefont {Ghosh}}, \bibinfo {author} {\bibfnamefont {S.}~\bibnamefont {Yang}}, \bibinfo {author} {\bibfnamefont {Y.}~\bibnamefont {Dai}}, \bibinfo {author} {\bibfnamefont {Z.}~\bibnamefont {Zhou}}, \bibinfo {author} {\bibfnamefont {T.}~\bibnamefont {Wang}}, \bibinfo {author} {\bibfnamefont {C.-B.}\ \bibnamefont {Huang}},\ and\ \bibinfo {author} {\bibfnamefont {H.}~\bibnamefont {Petek}},\ }\bibfield  {title} {\bibinfo {title} {A topological lattice of plasmonic merons},\ }\href@noop {} {\bibfield  {journal} {\bibinfo  {journal} {Applied Physics Reviews}\ }\textbf {\bibinfo {volume} {8}} (\bibinfo {year} {2021})}\BibitemShut {NoStop}%
\bibitem [{\citenamefont {Ghosh}\ \emph {et~al.}(2023)\citenamefont {Ghosh}, \citenamefont {Yang}, \citenamefont {Dai},\ and\ \citenamefont {Petek}}]{spin_merons_polygons}%
  \BibitemOpen
  \bibfield  {author} {\bibinfo {author} {\bibfnamefont {A.}~\bibnamefont {Ghosh}}, \bibinfo {author} {\bibfnamefont {S.}~\bibnamefont {Yang}}, \bibinfo {author} {\bibfnamefont {Y.}~\bibnamefont {Dai}},\ and\ \bibinfo {author} {\bibfnamefont {H.}~\bibnamefont {Petek}},\ }\bibfield  {title} {\bibinfo {title} {The spin texture topology of polygonal plasmon fields},\ }\href@noop {} {\bibfield  {journal} {\bibinfo  {journal} {ACS Photonics}\ }\textbf {\bibinfo {volume} {10}},\ \bibinfo {pages} {13} (\bibinfo {year} {2023})}\BibitemShut {NoStop}%
\bibitem [{\citenamefont {Ber{\v{s}}kys}\ and\ \citenamefont {Orlov}(2023)}]{Airy_beams_merons}%
  \BibitemOpen
  \bibfield  {author} {\bibinfo {author} {\bibfnamefont {J.}~\bibnamefont {Ber{\v{s}}kys}}\ and\ \bibinfo {author} {\bibfnamefont {S.}~\bibnamefont {Orlov}},\ }\bibfield  {title} {\bibinfo {title} {Accelerating {A}iry beams with particle-like polarization topologies and free-space bimeronic lattices},\ }\href@noop {} {\bibfield  {journal} {\bibinfo  {journal} {Optics Letters}\ }\textbf {\bibinfo {volume} {48}},\ \bibinfo {pages} {1168} (\bibinfo {year} {2023})}\BibitemShut {NoStop}%
\bibitem [{\citenamefont {Shen}(2021)}]{Shen_bimerons}%
  \BibitemOpen
  \bibfield  {author} {\bibinfo {author} {\bibfnamefont {Y.}~\bibnamefont {Shen}},\ }\bibfield  {title} {\bibinfo {title} {Topological bimeronic beams},\ }\href@noop {} {\bibfield  {journal} {\bibinfo  {journal} {Optics Letters}\ }\textbf {\bibinfo {volume} {46}},\ \bibinfo {pages} {3737} (\bibinfo {year} {2021})}\BibitemShut {NoStop}%
\bibitem [{\citenamefont {Shen}\ \emph {et~al.}(2024{\natexlab{b}})\citenamefont {Shen}, \citenamefont {He}, \citenamefont {Song}, \citenamefont {Chen}, \citenamefont {He}, \citenamefont {Ma}, \citenamefont {Fells}, \citenamefont {Elston}, \citenamefont {Morris}, \citenamefont {Booth},\ and\ \citenamefont {Forbes}}]{skyrmions_lens}%
  \BibitemOpen
  \bibfield  {author} {\bibinfo {author} {\bibfnamefont {Y.}~\bibnamefont {Shen}}, \bibinfo {author} {\bibfnamefont {C.}~\bibnamefont {He}}, \bibinfo {author} {\bibfnamefont {Z.}~\bibnamefont {Song}}, \bibinfo {author} {\bibfnamefont {B.}~\bibnamefont {Chen}}, \bibinfo {author} {\bibfnamefont {H.}~\bibnamefont {He}}, \bibinfo {author} {\bibfnamefont {Y.}~\bibnamefont {Ma}}, \bibinfo {author} {\bibfnamefont {J.~A.}\ \bibnamefont {Fells}}, \bibinfo {author} {\bibfnamefont {S.~J.}\ \bibnamefont {Elston}}, \bibinfo {author} {\bibfnamefont {S.~M.}\ \bibnamefont {Morris}}, \bibinfo {author} {\bibfnamefont {M.~J.}\ \bibnamefont {Booth}},\ and\ \bibinfo {author} {\bibfnamefont {A.}~\bibnamefont {Forbes}},\ }\bibfield  {title} {\bibinfo {title} {Topologically controlled multiskyrmions in photonic gradient-index lenses},\ }\href@noop {} {\bibfield  {journal} {\bibinfo  {journal} {Physical Review Applied}\ }\textbf {\bibinfo {volume} {21}},\ \bibinfo {pages} {024025} (\bibinfo {year} {2024}{\natexlab{b}})}\BibitemShut
  {NoStop}%
\bibitem [{\citenamefont {Beckley}\ \emph {et~al.}(2010)\citenamefont {Beckley}, \citenamefont {Brown},\ and\ \citenamefont {Alonso}}]{full_Poincare_beams}%
  \BibitemOpen
  \bibfield  {author} {\bibinfo {author} {\bibfnamefont {A.~M.}\ \bibnamefont {Beckley}}, \bibinfo {author} {\bibfnamefont {T.~G.}\ \bibnamefont {Brown}},\ and\ \bibinfo {author} {\bibfnamefont {M.~A.}\ \bibnamefont {Alonso}},\ }\bibfield  {title} {\bibinfo {title} {Full {P}oincar{\'e} beams},\ }\href@noop {} {\bibfield  {journal} {\bibinfo  {journal} {Optics Express}\ }\textbf {\bibinfo {volume} {18}},\ \bibinfo {pages} {10777} (\bibinfo {year} {2010})}\BibitemShut {NoStop}%
\bibitem [{\citenamefont {Donati}\ \emph {et~al.}(2016)\citenamefont {Donati}, \citenamefont {Dominici}, \citenamefont {Dagvadorj}, \citenamefont {Ballarini}, \citenamefont {De~Giorgi}, \citenamefont {Bramati}, \citenamefont {Gigli}, \citenamefont {Rubo}, \citenamefont {Szyma{\'n}ska},\ and\ \citenamefont {Sanvitto}}]{Poincare_skyrmions}%
  \BibitemOpen
  \bibfield  {author} {\bibinfo {author} {\bibfnamefont {S.}~\bibnamefont {Donati}}, \bibinfo {author} {\bibfnamefont {L.}~\bibnamefont {Dominici}}, \bibinfo {author} {\bibfnamefont {G.}~\bibnamefont {Dagvadorj}}, \bibinfo {author} {\bibfnamefont {D.}~\bibnamefont {Ballarini}}, \bibinfo {author} {\bibfnamefont {M.}~\bibnamefont {De~Giorgi}}, \bibinfo {author} {\bibfnamefont {A.}~\bibnamefont {Bramati}}, \bibinfo {author} {\bibfnamefont {G.}~\bibnamefont {Gigli}}, \bibinfo {author} {\bibfnamefont {Y.~G.}\ \bibnamefont {Rubo}}, \bibinfo {author} {\bibfnamefont {M.~H.}\ \bibnamefont {Szyma{\'n}ska}},\ and\ \bibinfo {author} {\bibfnamefont {D.}~\bibnamefont {Sanvitto}},\ }\bibfield  {title} {\bibinfo {title} {Twist of generalized skyrmions and spin vortices in a polariton superfluid},\ }\href@noop {} {\bibfield  {journal} {\bibinfo  {journal} {Proceedings of the National Academy of Sciences}\ }\textbf {\bibinfo {volume} {113}},\ \bibinfo {pages} {14926} (\bibinfo {year} {2016})}\BibitemShut {NoStop}%
\bibitem [{\citenamefont {Gao}\ \emph {et~al.}(2020)\citenamefont {Gao}, \citenamefont {Speirits}, \citenamefont {Castellucci}, \citenamefont {Franke-Arnold}, \citenamefont {Barnett},\ and\ \citenamefont {G{\"o}tte}}]{paraxial_skyrmionic_beams}%
  \BibitemOpen
  \bibfield  {author} {\bibinfo {author} {\bibfnamefont {S.}~\bibnamefont {Gao}}, \bibinfo {author} {\bibfnamefont {F.~C.}\ \bibnamefont {Speirits}}, \bibinfo {author} {\bibfnamefont {F.}~\bibnamefont {Castellucci}}, \bibinfo {author} {\bibfnamefont {S.}~\bibnamefont {Franke-Arnold}}, \bibinfo {author} {\bibfnamefont {S.~M.}\ \bibnamefont {Barnett}},\ and\ \bibinfo {author} {\bibfnamefont {J.~B.}\ \bibnamefont {G{\"o}tte}},\ }\bibfield  {title} {\bibinfo {title} {Paraxial skyrmionic beams},\ }\href@noop {} {\bibfield  {journal} {\bibinfo  {journal} {Physical Review A}\ }\textbf {\bibinfo {volume} {102}},\ \bibinfo {pages} {053513} (\bibinfo {year} {2020})}\BibitemShut {NoStop}%
\bibitem [{\citenamefont {Pal}\ and\ \citenamefont {Senthilkumaran}(2016)}]{lemon_fields}%
  \BibitemOpen
  \bibfield  {author} {\bibinfo {author} {\bibfnamefont {S.~K.}\ \bibnamefont {Pal}}\ and\ \bibinfo {author} {\bibfnamefont {P.}~\bibnamefont {Senthilkumaran}},\ }\bibfield  {title} {\bibinfo {title} {Cultivation of lemon fields},\ }\href@noop {} {\bibfield  {journal} {\bibinfo  {journal} {Optics Express}\ }\textbf {\bibinfo {volume} {24}},\ \bibinfo {pages} {28008} (\bibinfo {year} {2016})}\BibitemShut {NoStop}%
\bibitem [{\citenamefont {Marco}\ \emph {et~al.}(2024)\citenamefont {Marco}, \citenamefont {Herrera}, \citenamefont {Brasselet},\ and\ \citenamefont {Alonso}}]{prop_invariant_meron_lattices}%
  \BibitemOpen
  \bibfield  {author} {\bibinfo {author} {\bibfnamefont {D.}~\bibnamefont {Marco}}, \bibinfo {author} {\bibfnamefont {I.}~\bibnamefont {Herrera}}, \bibinfo {author} {\bibfnamefont {S.}~\bibnamefont {Brasselet}},\ and\ \bibinfo {author} {\bibfnamefont {M.~A.}\ \bibnamefont {Alonso}},\ }\bibfield  {title} {\bibinfo {title} {Propagation-invariant optical meron lattices},\ }\href@noop {} {\bibfield  {journal} {\bibinfo  {journal} {ACS Photonics}\ }\textbf {\bibinfo {volume} {11}},\ \bibinfo {pages} {2397} (\bibinfo {year} {2024})}\BibitemShut {NoStop}%
\bibitem [{\citenamefont {Fathauer}(2020)}]{Fathauer_tessellations}%
  \BibitemOpen
  \bibfield  {author} {\bibinfo {author} {\bibfnamefont {R.}~\bibnamefont {Fathauer}},\ }\href@noop {} {\emph {\bibinfo {title} {Tessellations: mathematics, art, and recreation}}}\ (\bibinfo  {publisher} {CRC Press},\ \bibinfo {year} {2020})\ p.~\bibinfo {pages} {21}\BibitemShut {NoStop}%
\bibitem [{\citenamefont {Peirce}(1879)}]{Peirce}%
  \BibitemOpen
  \bibfield  {author} {\bibinfo {author} {\bibfnamefont {C.~S.}\ \bibnamefont {Peirce}},\ }\bibfield  {title} {\bibinfo {title} {A quincuncial projection of the sphere},\ }\href@noop {} {\bibfield  {journal} {\bibinfo  {journal} {American Journal of Mathematics}\ }\textbf {\bibinfo {volume} {2}},\ \bibinfo {pages} {394} (\bibinfo {year} {1879})}\BibitemShut {NoStop}%
\bibitem [{\citenamefont {Adams}(1925)}]{Adams_hexagon}%
  \BibitemOpen
  \bibfield  {author} {\bibinfo {author} {\bibfnamefont {O.~S.}\ \bibnamefont {Adams}},\ }\href@noop {} {\emph {\bibinfo {title} {Elliptic functions applied to conformal world maps}}},\ Vol.\ \bibinfo {volume} {297}\ (\bibinfo  {publisher} {US Government Printing Office},\ \bibinfo {year} {1925})\ pp.\ \bibinfo {pages} {72--78}\BibitemShut {NoStop}%
\bibitem [{\citenamefont {Lee}(1965)}]{Lee_tetrahedron}%
  \BibitemOpen
  \bibfield  {author} {\bibinfo {author} {\bibfnamefont {L.}~\bibnamefont {Lee}},\ }\bibfield  {title} {\bibinfo {title} {Some conformal projections based on elliptic functions},\ }\href@noop {} {\bibfield  {journal} {\bibinfo  {journal} {Geographical Review}\ }\textbf {\bibinfo {volume} {55}},\ \bibinfo {pages} {563} (\bibinfo {year} {1965})}\BibitemShut {NoStop}%
\bibitem [{\citenamefont {Lee}(1976{\natexlab{a}})}]{Lee_Dixon}%
  \BibitemOpen
  \bibfield  {author} {\bibinfo {author} {\bibfnamefont {L.}~\bibnamefont {Lee}},\ }\bibfield  {title} {\bibinfo {title} {Conformal projections based on {D}ixon elliptic functions},\ }\href@noop {} {\bibfield  {journal} {\bibinfo  {journal} {Cartographica: The International Journal for Geographic Information and Geovisualization}\ }\textbf {\bibinfo {volume} {13}},\ \bibinfo {pages} {38} (\bibinfo {year} {1976}{\natexlab{a}})}\BibitemShut {NoStop}%
\bibitem [{\citenamefont {Schwarz}(1869)}]{Schwarz}%
  \BibitemOpen
  \bibfield  {author} {\bibinfo {author} {\bibfnamefont {H.~A.}\ \bibnamefont {Schwarz}},\ }\bibfield  {title} {\bibinfo {title} {Ueber einige abbildungsaufgaben},\ }\href@noop {} {\bibfield  {journal} {\bibinfo  {journal} {Journal für die reine und angewandte Mathematik}\ }\textbf {\bibinfo {volume} {13}},\ \bibinfo {pages} {105} (\bibinfo {year} {1869})}\BibitemShut {NoStop}%
\bibitem [{\citenamefont {Lee}(1976{\natexlab{b}})}]{Lee_Jacobi}%
  \BibitemOpen
  \bibfield  {author} {\bibinfo {author} {\bibfnamefont {L.}~\bibnamefont {Lee}},\ }\bibfield  {title} {\bibinfo {title} {Conformal projections based on {J}acobian elliptic functions},\ }\href@noop {} {\bibfield  {journal} {\bibinfo  {journal} {Cartographica: The International Journal for Geographic Information and Geovisualization}\ }\textbf {\bibinfo {volume} {13}},\ \bibinfo {pages} {67} (\bibinfo {year} {1976}{\natexlab{b}})}\BibitemShut {NoStop}%
\bibitem [{\citenamefont {van Fossen~Conrad}\ and\ \citenamefont {Flajolet}(2006)}]{elliptic_functions_Conrad}%
  \BibitemOpen
  \bibfield  {author} {\bibinfo {author} {\bibfnamefont {E.}~\bibnamefont {van Fossen~Conrad}}\ and\ \bibinfo {author} {\bibfnamefont {P.}~\bibnamefont {Flajolet}},\ }\bibfield  {title} {\bibinfo {title} {The {F}ermat cubic, elliptic functions, continued fractions, and a combinatorial excursion},\ }\href@noop {} {\bibfield  {journal} {\bibinfo  {journal} {S{\'e}minaire Lotharingien de Combinatoire}\ }\textbf {\bibinfo {volume} {54}},\ \bibinfo {pages} {1} (\bibinfo {year} {2006})}\BibitemShut {NoStop}%
\bibitem [{\citenamefont {Freund}(2004)}]{Freund_lattices}%
  \BibitemOpen
  \bibfield  {author} {\bibinfo {author} {\bibfnamefont {I.}~\bibnamefont {Freund}},\ }\bibfield  {title} {\bibinfo {title} {Polarization singularities in optical lattices},\ }\href@noop {} {\bibfield  {journal} {\bibinfo  {journal} {Optics Letters}\ }\textbf {\bibinfo {volume} {29}},\ \bibinfo {pages} {875} (\bibinfo {year} {2004})}\BibitemShut {NoStop}%
\bibitem [{\citenamefont {Dennis}(2002)}]{Dennis_singularities}%
  \BibitemOpen
  \bibfield  {author} {\bibinfo {author} {\bibfnamefont {M.}~\bibnamefont {Dennis}},\ }\bibfield  {title} {\bibinfo {title} {Polarization singularities in paraxial vector fields: morphology and statistics},\ }\href@noop {} {\bibfield  {journal} {\bibinfo  {journal} {Optics Communications}\ }\textbf {\bibinfo {volume} {213}},\ \bibinfo {pages} {201} (\bibinfo {year} {2002})}\BibitemShut {NoStop}%
\bibitem [{\citenamefont {Maurer}\ \emph {et~al.}(2007)\citenamefont {Maurer}, \citenamefont {Jesacher}, \citenamefont {F{\"u}rhapter}, \citenamefont {Bernet},\ and\ \citenamefont {Ritsch-Marte}}]{vector_setup}%
  \BibitemOpen
  \bibfield  {author} {\bibinfo {author} {\bibfnamefont {C.}~\bibnamefont {Maurer}}, \bibinfo {author} {\bibfnamefont {A.}~\bibnamefont {Jesacher}}, \bibinfo {author} {\bibfnamefont {S.}~\bibnamefont {F{\"u}rhapter}}, \bibinfo {author} {\bibfnamefont {S.}~\bibnamefont {Bernet}},\ and\ \bibinfo {author} {\bibfnamefont {M.}~\bibnamefont {Ritsch-Marte}},\ }\bibfield  {title} {\bibinfo {title} {Tailoring of arbitrary optical vector beams},\ }\href@noop {} {\bibfield  {journal} {\bibinfo  {journal} {New Journal of Physics}\ }\textbf {\bibinfo {volume} {9}},\ \bibinfo {pages} {78} (\bibinfo {year} {2007})}\BibitemShut {NoStop}%
\bibitem [{\citenamefont {Bolduc}\ \emph {et~al.}(2013)\citenamefont {Bolduc}, \citenamefont {Bent}, \citenamefont {Santamato}, \citenamefont {Karimi},\ and\ \citenamefont {Boyd}}]{method_modulation}%
  \BibitemOpen
  \bibfield  {author} {\bibinfo {author} {\bibfnamefont {E.}~\bibnamefont {Bolduc}}, \bibinfo {author} {\bibfnamefont {N.}~\bibnamefont {Bent}}, \bibinfo {author} {\bibfnamefont {E.}~\bibnamefont {Santamato}}, \bibinfo {author} {\bibfnamefont {E.}~\bibnamefont {Karimi}},\ and\ \bibinfo {author} {\bibfnamefont {R.~W.}\ \bibnamefont {Boyd}},\ }\bibfield  {title} {\bibinfo {title} {Exact solution to simultaneous intensity and phase encryption with a single phase-only hologram},\ }\href@noop {} {\bibfield  {journal} {\bibinfo  {journal} {Optics Letters}\ }\textbf {\bibinfo {volume} {38}},\ \bibinfo {pages} {3546} (\bibinfo {year} {2013})}\BibitemShut {NoStop}%
\bibitem [{\citenamefont {Lee}(1976{\natexlab{c}})}]{Lee_circle}%
  \BibitemOpen
  \bibfield  {author} {\bibinfo {author} {\bibfnamefont {L.}~\bibnamefont {Lee}},\ }\bibfield  {title} {\bibinfo {title} {Conformal projections of the sphere or parts of the sphere within a circle},\ }\href@noop {} {\bibfield  {journal} {\bibinfo  {journal} {Cartographica: The International Journal for Geographic Information and Geovisualization}\ }\textbf {\bibinfo {volume} {13}},\ \bibinfo {pages} {15} (\bibinfo {year} {1976}{\natexlab{c}})}\BibitemShut {NoStop}%
\bibitem [{\citenamefont {Adams}(1929)}]{Adams_square}%
  \BibitemOpen
  \bibfield  {author} {\bibinfo {author} {\bibfnamefont {O.~S.}\ \bibnamefont {Adams}},\ }\href@noop {} {\emph {\bibinfo {title} {Conformal projection of the sphere within a square}}},\ Vol.~\bibinfo {volume} {4}\ (\bibinfo  {publisher} {US Government Printing Office},\ \bibinfo {year} {1929})\BibitemShut {NoStop}%
\bibitem [{\citenamefont {Cox}(1936)}]{Cox}%
  \BibitemOpen
  \bibfield  {author} {\bibinfo {author} {\bibfnamefont {J.}~\bibnamefont {Cox}},\ }\bibfield  {title} {\bibinfo {title} {Repr{\'e}sentation de la surface enti{\`e}re de la terre dans un triangle {\'e}quilat{\'e}ral},\ }\href@noop {} {\bibfield  {journal} {\bibinfo  {journal} {Bulletin de la Classe des sciences. Acad{\'e}mie royale de Belgique}\ }\textbf {\bibinfo {volume} {21}},\ \bibinfo {pages} {66} (\bibinfo {year} {1936})}\BibitemShut {NoStop}%
\bibitem [{\citenamefont {Hansen}\ \emph {et~al.}(2016)\citenamefont {Hansen}, \citenamefont {Schultz},\ and\ \citenamefont {Bigelow}}]{hansen2016singular}%
  \BibitemOpen
  \bibfield  {author} {\bibinfo {author} {\bibfnamefont {A.}~\bibnamefont {Hansen}}, \bibinfo {author} {\bibfnamefont {J.~T.}\ \bibnamefont {Schultz}},\ and\ \bibinfo {author} {\bibfnamefont {N.~P.}\ \bibnamefont {Bigelow}},\ }\bibfield  {title} {\bibinfo {title} {Singular atom optics with spinor bose--einstein condensates},\ }\href@noop {} {\bibfield  {journal} {\bibinfo  {journal} {Optica}\ }\textbf {\bibinfo {volume} {3}},\ \bibinfo {pages} {355} (\bibinfo {year} {2016})}\BibitemShut {NoStop}%
\bibitem [{\citenamefont {Sugic}\ \emph {et~al.}(2021)\citenamefont {Sugic}, \citenamefont {Droop}, \citenamefont {Otte}, \citenamefont {Ehrmanntraut}, \citenamefont {Nori}, \citenamefont {Ruostekoski}, \citenamefont {Denz},\ and\ \citenamefont {Dennis}}]{sugic2021particle}%
  \BibitemOpen
  \bibfield  {author} {\bibinfo {author} {\bibfnamefont {D.}~\bibnamefont {Sugic}}, \bibinfo {author} {\bibfnamefont {R.}~\bibnamefont {Droop}}, \bibinfo {author} {\bibfnamefont {E.}~\bibnamefont {Otte}}, \bibinfo {author} {\bibfnamefont {D.}~\bibnamefont {Ehrmanntraut}}, \bibinfo {author} {\bibfnamefont {F.}~\bibnamefont {Nori}}, \bibinfo {author} {\bibfnamefont {J.}~\bibnamefont {Ruostekoski}}, \bibinfo {author} {\bibfnamefont {C.}~\bibnamefont {Denz}},\ and\ \bibinfo {author} {\bibfnamefont {M.~R.}\ \bibnamefont {Dennis}},\ }\bibfield  {title} {\bibinfo {title} {Particle-like topologies in light},\ }\href@noop {} {\bibfield  {journal} {\bibinfo  {journal} {Nature Communications}\ }\textbf {\bibinfo {volume} {12}},\ \bibinfo {pages} {6785} (\bibinfo {year} {2021})}\BibitemShut {NoStop}%
\bibitem [{\citenamefont {Marco}\ and\ \citenamefont {Alonso}(2022)}]{F3DP_fields}%
  \BibitemOpen
  \bibfield  {author} {\bibinfo {author} {\bibfnamefont {D.}~\bibnamefont {Marco}}\ and\ \bibinfo {author} {\bibfnamefont {M.~A.}\ \bibnamefont {Alonso}},\ }\bibfield  {title} {\bibinfo {title} {Optical fields spanning the 4{D} space of nonparaxial polarization},\ }\href@noop {} {\bibfield  {journal} {\bibinfo  {journal} {arXiv preprint arXiv:2212.01366}\ } (\bibinfo {year} {2022})}\BibitemShut {NoStop}%
\end{thebibliography}%

\newpage

\section*{Supplemental Information}

\subsection{Energy density and variational calculation}\label{sec:variational_calc}
Let us first write the conditions for a map ${\bf s}(x,y)$ to be conformal. Infinitesimal changes of the same size in each of the two flat coordinates $x,y$ must result in changes over the unit sphere of ${\bf s}$ also of equal size in two orthogonal directions. We can write this as
\begin{subequations}
\begin{align}
{\bf s}\times\partial_x{\bf s}&={\rm sgn}(\rho_{\rm S})\partial_y{\bf s},\\
{\bf s}\times\partial_y{\bf s}&=-{\rm sgn}(\rho_{\rm S})\partial_x{\bf s},
\end{align}
\end{subequations}
where ${\rm sgn}(\rho_{\rm S})$ is the sign of the Skyrme density (assumed to be constant), and the operation ${\bf s}\times$ performs a rotation of $90^\circ$ over the surface of the sphere. We can combine these equations as
\begin{align}
{\bf s}\times\nabla{\bf s}={\rm sgn}(\rho_{\rm S})\sigma_1\nabla{\bf s},
\end{align}
with
\begin{align}
\sigma_1=\left(\begin{array}{cc}0&1\\-1&0\end{array}\right).
\end{align}
That is, a rotation over the sphere corresponds to a rotation over the plane. 
The equation above can also be written as
\begin{align}
\nabla{\bf s}=-{\rm sgn}(\rho_{\rm S})\,{\bf s}\times(\sigma_1\nabla){\bf s},
\end{align}
where we have to be careful distinguishing the vector operations in the two-dimensional flat space and the three-dimensional ambient space for ${\bf s}$ (in this case the cross product). 
It is easy to see that
\begin{align}
\nabla^2{\bf s}=-{\rm sgn}(\rho_{\rm S})\left[(\nabla{\bf s})\times(\sigma_1\nabla{\bf s})+{\bf s}\times(\nabla\sigma_1\nabla){\bf s}\right]={\bf 0}.
\end{align}

Now let us consider a measure of stretching of ${\bf s}$ when distributed over a plane. This stretching can be considered as giving a local energy density given by the squared Euclidean norm of the derivative, that is 
\begin{align}
\epsilon=\frac12||\nabla{\bf s}||^2=\frac12(\partial_is_j)(\partial_is_j),
\end{align}
where we used Einstein's convention of implicit sum over repeated indices, with $i=x,y$ and $j=1,2,3$. The total energy over a unit cell $\sigma$ is then given by
\begin{align}
\mu=\int_\sigma\epsilon\,{\rm d}x{\rm d}y.
\end{align}
We now consider the functional derivative with respect to the function ${\bf s}$ of this energy. By using integration by parts to remove the derivatives of each factor of ${\bf s}$ and realizing that the integrated terms cancel due to the periodicity of the pattern, we arrive at
\begin{align}
\frac{\delta}{\delta{\bf s}}\mu=-\nabla^2{\bf s}.
\end{align}
This variation is then zero for a conformal distribution.

Note that for a conformal map $\partial_x{\bf s}$ and $\partial_y{\bf s}$ have the same magnitude and are perpendicular, so the magnitude of the Skyrme density becomes
\begin{align}
|\rho_{\rm S}|&=\left|{\bf s}\cdot\partial_x{\bf s}\times\partial_y{\bf s}\right|\nonumber\\
&=|\partial_x{\bf s}||\partial_y{\bf s}|\nonumber\\
&=\frac{|\partial_x{\bf s}|^2+|\partial_y{\bf s}|^2}2\nonumber\\
&=\epsilon,
\end{align}
So for all conformal maps $\mu=4\pi |N_{\rm S}|$. 

\subsection{Inverse mappings}\label{sec:inverse_mappings}

We provide here the inverse maps of the transformations $w_i(z_i)$ 
in Eqs.~\eqref{eq:maps} in the main text, where $i=\mathrm{P},\mathrm{A},\mathrm{L},\mathrm{W}$ stand for the Peirce (P), Adams (A), Lee (L) and Wray (W) projections:
\begin{subequations}
\begin{align}
    z_\mathrm{P}(w_\mathrm{P})&=z_0+ \frac{\sqrt2 d}{\gamma_1} e^{\im \alpha} F\left(\arcsin\tilde{w}_\mathrm{P} ,\im\right), \label{eq:inverse_Peirce}\\
    z_\mathrm{A}(w_\mathrm{A})&=z_0+ \frac{d}{\gamma_2} \tilde{w}_\mathrm{A} e^{\im \alpha} {}_2 F_1\left(\frac{1}{3},\frac{2}{3};\frac{4}{3};\tilde{w}^3_\mathrm{A}\right),\label{eq:inverse_Adams}\\
    z_\mathrm{L}(w_\mathrm{L})&=z_0+ \frac{d}{\gamma_2} f(\tilde{w}_\mathrm{L}) e^{\im \alpha} {}_2 F_1\left[\frac{1}{3},\frac{2}{3};\frac{4}{3};f^3(\tilde{w}_\mathrm{L})\right],\label{eq:inverse_Lee}\\
    z_\mathrm{W}(w_\mathrm{W})&=z_0+ \frac{d}{\gamma_3}e^{\im \alpha} \mathrm{sd}^{-1}\left(2\tilde{w}_\mathrm{W} ,\frac{\sqrt3-1}{2\sqrt2}\right),\label{eq:inverse_Wray}
 \end{align}
 \end{subequations}
with
\begin{align}
    f^3(\tilde{w}_\mathrm{L})=\frac{1}{2}\left(1-\sqrt{1-2\sqrt2\tilde{w}^3_\mathrm{L}}\right),
\end{align}
and
\begin{align}
    \tilde{w}_{\mathrm{P},\mathrm{A},\mathrm{L},\mathrm{W}}=e^{-\im (\phi_0+\alpha)} w_{\mathrm{P},\mathrm{A},\mathrm{L},\mathrm{W}},
\end{align}
where $F$ represents the incomplete elliptic integral of the first kind, ${}_2 F_1$ stands for the ordinary hypergeometric function, and $\mathrm{sd}^{-1}$ denotes the inverse of the Jacobi elliptic function sd, and the
parameters $\gamma_{1,2,3}$ are given in the main text, as is a description of 
the effect of the parameters $\alpha$, $\phi_0$, $z_0$ and $d$.

\begin{figure}
\centering
\def\svgwidth{0.44\textwidth}
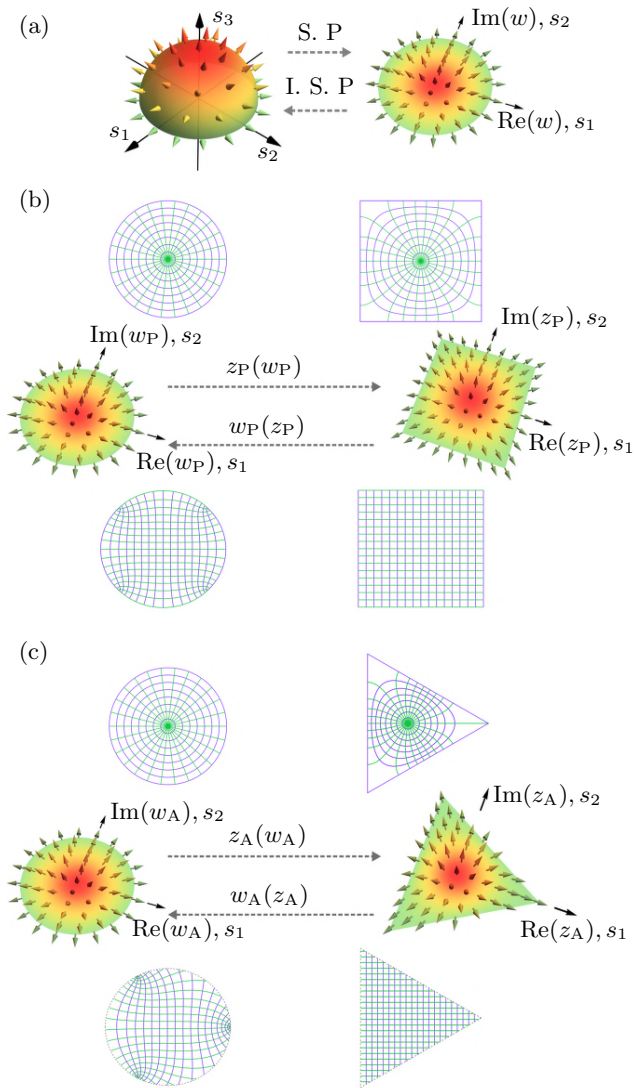\caption{\label{fig:Schwarz_maps} (a) Depiction of the stereographic projection (S. P) and of the inverse stereographic projection (I. S. P) between a meron within the unit disk and the unit sphere's 
northern hemisphere. (b,c) Schwarz maps and their inverse transformations between a meron within the unit disk and a (b) squared and a (c) equilateral triangular meron.}
\end{figure}

Equations \eqref{eq:maps} in the main text define complex quantities, $w_i(z_i)$, which, via an inverse stereographic projection, yield values for the azimuthal and polar spherical angles, $\phi=\mathrm{arg}\left[w_i(z_i)\right]$ and $\theta=2\arctan |w_i(z_i)|$, respectively, for any point on the plane. The $x$ and $y$ Cartesian coordinates of this point correspond to the real and imaginary parts of the complex number $z_i$, respectively. On the other hand, the real and imaginary parts of the inverse transformations $z_i(w_i)$ provide a pair of Cartesian coordinates on the plane where a specific value of $\phi$ and $\theta$ is mapped. The spherical coordinates are related to $w_i$ through a stereographic projection as $w_i= \tan(\theta/2) e^{\im\phi}$.

Equations \eqref{eq:inverse_Peirce} and \eqref{eq:inverse_Adams} are well-known Schwarz maps. The interior of the unit disk can be mapped conformally onto the interior of a regular polygon of $n$ sides through the Schwarz integral \cite{Schwarz}:
\begin{align}
 z=\int_{0}^{w} \frac{1}{\left(1-w^n\right)^{2/n}}\mathrm{d}w,
\end{align}
where the real and imaginary parts of $w$ are the Cartesian coordinates within the disk, and the real and imaginary parts of $z$ are the Cartesian coordinates within the polygon. The first two equations result from this integral for a square ($n=4$) and for an equilateral triangle ($n=3$) \cite{elliptic_functions_Conrad}, but we introduced the additional parameters $\alpha$, $\phi_0$, $z_0$ and $d$ to control certain properties of the map, as explained in the main text.

Figure \ref{fig:Schwarz_maps}(a) illustrates the transformation between a meron within the unit disk and the northern hemisphere of the unit sphere through the stereographic projection and its inverse. Figures \ref{fig:Schwarz_maps}(b,c) illustrate the Schwarz maps, $z_i(w_i)$, and their inverse functions, $w_i(z_i)$, between a meron within the unit disk, and a meron within a square ($i=\mathrm{P}$) and within an equilateral triangle ($i=\mathrm{A}$), respectively. When these maps are extended to the whole plane, they yield the entire Peirce and Adams textures shown in Fig.~\ref{fig:cartography_skyrmions} of the main document.

The inverse map of $w_\mathrm{L}(z_\mathrm{L})$ (namely $z_\mathrm{L}(w_\mathrm{L})$ in Eq.~\eqref{eq:inverse_Lee}) was derived by expressing the Dixon elliptic function $\mathrm{cm}z$ in the transformation $w_\mathrm{L}(z_\mathrm{L})$ (Eq.~\eqref{eq:Lee} in the main text) in terms of the Dixon elliptic function $\mathrm{sm}z$ using the fact that $\textrm{cm}^3 z + \textrm{sm}^3 z=1$. We then use Eq.~\eqref{eq:inverse_Adams}, which is the inverse function of $\mathrm{sm}z$ up to certain multiplicative and additive factors, to calculate the inverse of $w_\mathrm{L}(z_\mathrm{L})$.

\subsection{Additional theoretical distributions of the textures and fields}\label{sec:additional_theory}

\begin{figure}
    \centering
    \def\svgwidth{0.48\textwidth}
    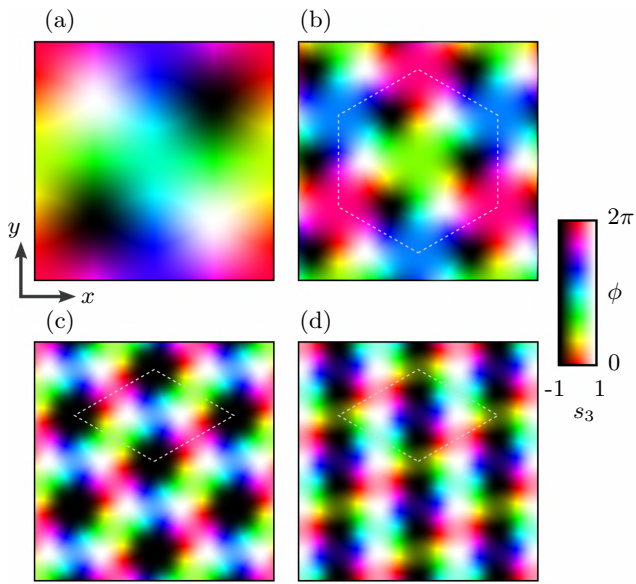\caption{\label{fig:hue_brightness_textures} Distributions of $\phi$, $s_3$ for the textures resulting from (a) Peirce's, (b) Adams', (c) Lee's and (d) Wray's projections. The plot shown in (a) constitutes the entire unit cell of Peirce's texture. The unit cells in (b-d) are delimited by dashed lines.
    }
    \end{figure}

\begin{figure}
    \centering
    \def\svgwidth{0.48\textwidth}
    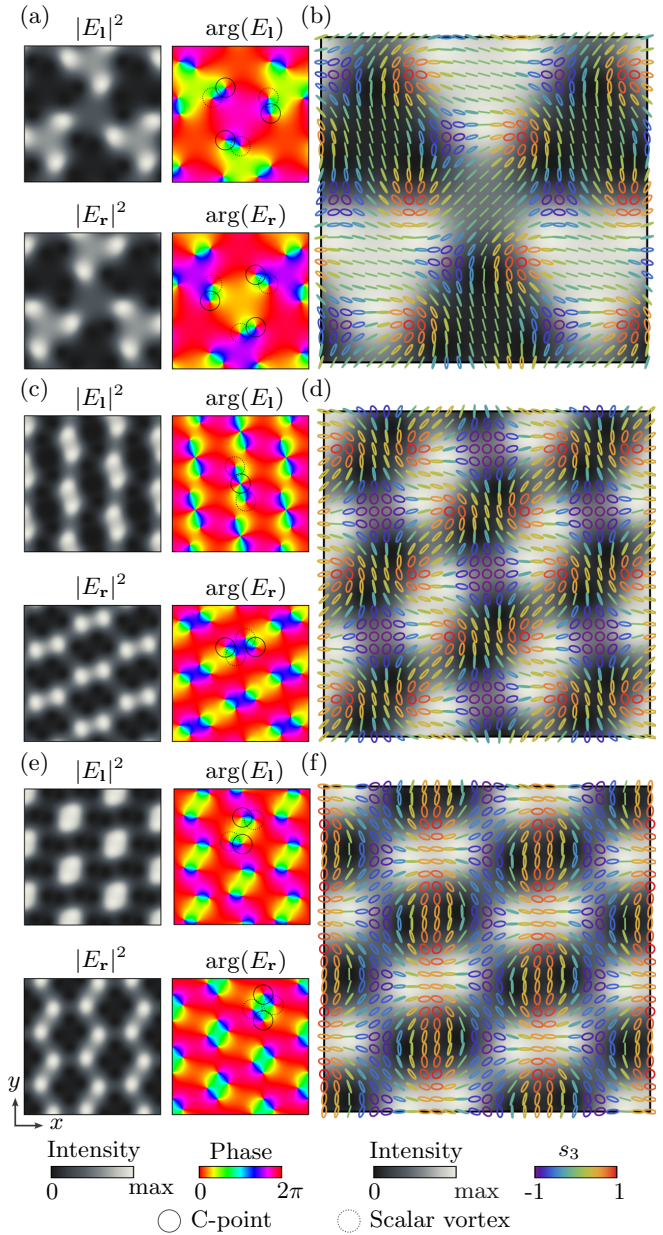\caption{\label{fig:fields_supplemental} (a,c,e) Theoretical intensity and phase distributions of the circular polarization components, and (b,d,f) polarization and total intensity distributions, for (a,b) Adams, (c,d) Lee and (e,f) Wray fields. We indicate which vortices in the circular components render either C-points or scalar vortices.
    }
    \end{figure}

\begin{figure}
    \centering
    \def\svgwidth{0.46\textwidth}
    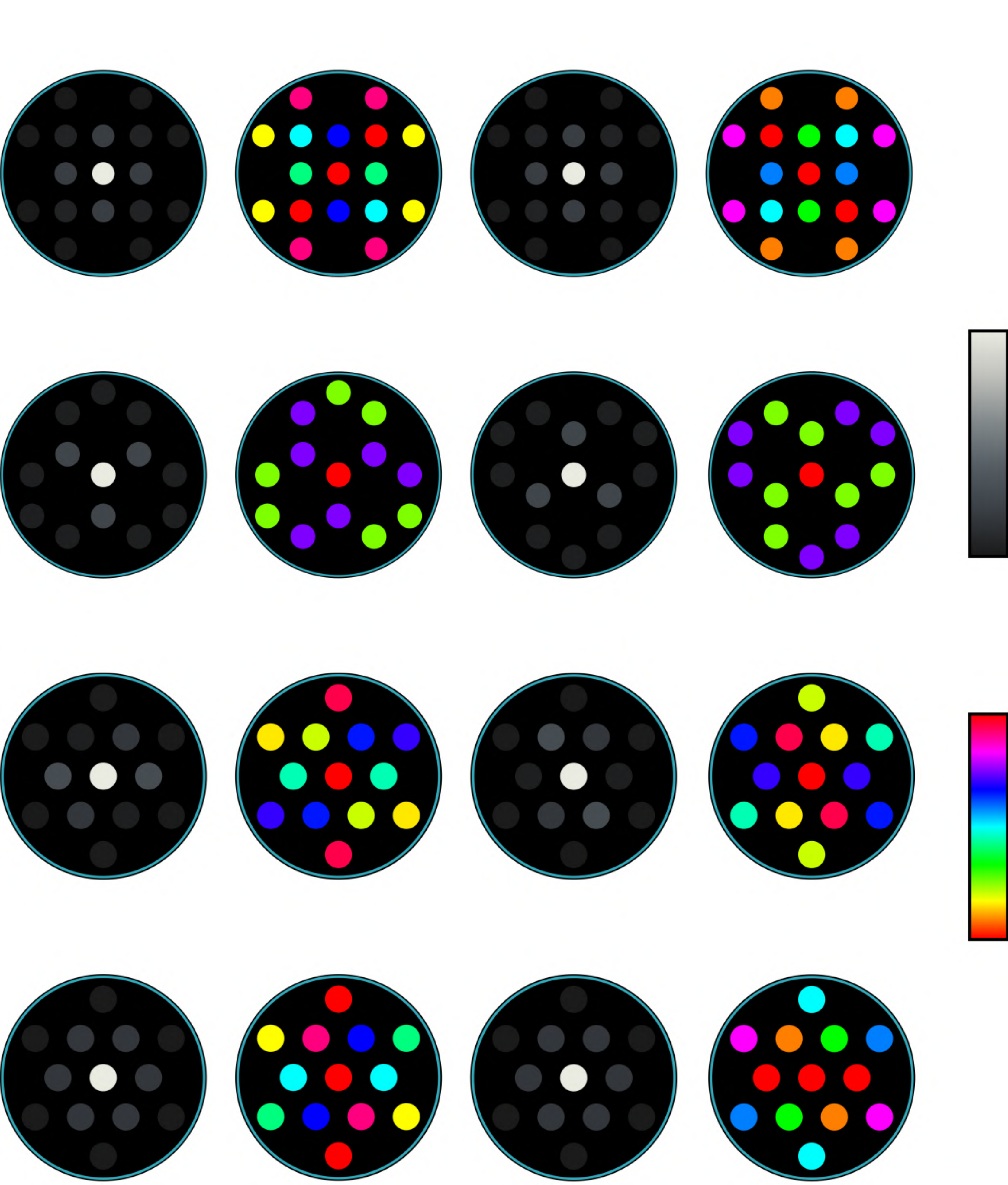\caption{\label{fig:spectrums} Intensity and phase of the first diffraction orders of the spatial spectrum of the circular polarization components of (a) Peirce, (b) Adams, (c) Lee and (d) Wray fields.
    }
    \end{figure}

Here, we show additional results for the textures and present the optical fields implementing the Adams, Lee and Wray textures. The parameters in Eqs.~\eqref{eq:maps} in the main text are set as 
follows. 
For Adams' projection: $\alpha=0$, $\phi_0=\pi/2$ and $z_0=-1/\sqrt3$. For Lee's projection: $\alpha=0$, $\phi_0=-5\pi/6$ and $z_0=1/\sqrt3$. For Wray's projection: $\alpha=\pi/2$, $\phi_0=-7\pi/12$ and $z_0=\sqrt3/4$.
We show a representation of the parameters $\phi$ and $s_3$ for every field in Fig.~\ref{fig:hue_brightness_textures}.
    The intensity and phase distributions of the circular polarization components, $E_{\mathbf{l},\mathbf{r}}$, as well as the polarization distributions, are presented in Fig.~\ref{fig:fields_supplemental} for Adams, Lee and Wray fields.
    As pointed out in the main text, the fields display a discrete spatial Fourier spectrum. The intensity and phase of the first diffraction orders of the spectrum for each field, which contain most of the energy, are presented in Fig.~\ref{fig:spectrums}. 

\subsection{Topological implications of Skyrme density sign preservation in skyrmionic textures}\label{sec:topological_properties}

In the main text, a mathematical result was utilized to demonstrate that all paraxial optical fields implementing periodic skyrmionic Stokes textures preserving the sign of the Skyrme density, $\mathrm{sgn}(\rho_\mathrm{S})$, exhibit zeros. The employed result establishes that for a Stokes texture with $\rho_\mathrm{S} > 0$ throughout space, all left(right)-handed C-points arise from a vortex in the relative phase between the circular polarization components, denoted as $\phi$, having a positive (negative) topological charge, $m_\phi$. Consequently, it follows that all C-points with $|m_\phi| = 1$ present left-handed lemons and right-handed stars. It is worth noting that the forthcoming result we demonstrate can be extended to fields characterized by a negative $\mathrm{sgn}(\rho_\mathrm{S})$ across the entire space, where the scenario would be reversed.

Consider mapping a small patch of the sphere. Let the polar coordinates of the local planar map be $(r,\varphi)$. When mapping the north pole, $r\approx\kappa\theta,\varphi=\phi$, where $\kappa$ is a positive constant. The Skyrme density has the same sign as the Jacobian between $(\theta,\phi)$ and $(r,\varphi)$, which equals $1/\kappa>0$. Now consider transporting continuously the map to the south pole without reversing local directions (i.e., $\rho_\mathrm{S}$ does not change sign). At the south pole the coordinates can be set to $r\approx\kappa(\pi-\theta),\varphi=-\phi$, which yield the same positive Jacobian. However, $m_\phi=\partial_\varphi\phi$ is positive at the north pole (left-circular polarization) and negative at the south pole (right-circular polarization).

\subsection{Experimental setup}\label{sec:experiment}

\begin{figure}
    \centering
    \def\svgwidth{0.48\textwidth}
    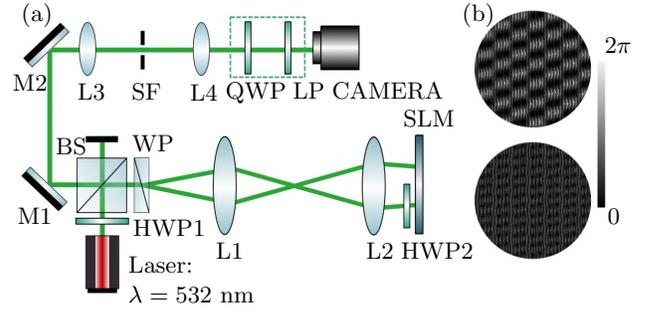\caption{\label{fig:experiment} (a) Experimental setup (not to scale). A linearly polarized continuous-wave (CW) laser beam ($\lambda=532$ nm) passes through a half-wave plate (HWP1) that controls the direction of polarization. The beam is then redirected by a 50:50 beamsplitter (BS) to a Wollaston prism (WP) that separates it into two orthogonal linearly polarized beams by 1$^{\circ}$. The two beams are magnified by a telescope composed of lenses L1 and L2 with focal lengths of $f_1 = 100$ mm and $f_2=500$ mm, respectively. Due to the fact that the SLM can only modulate the horizontal polarization, a HWP2 rotates the vertical polarization by 90$^{\circ}$. When the two beams are reflected by the SLM, the HWP2 rotates by 90$^{\circ}$ the beam polarization again. The two orthogonal polarized beams are then recombined by the WP and are separated from the initial beam by the BS. With a second telescope (composed of lenses L3 and L4 with focal lengths of $f_3 =300 $ mm and $f_4=400 $ mm, respectively) the SLM plane is imaged onto a CMOS camera. A spatial filter (SF) is placed at the Fourier plane of L4. In order to measure the polarization projections, a retractable quarter wave-plate (QWP) and a retractable linear polarizer (LP) are placed in front of the camera (enclosed within dashed lines). (b) Example of two holograms displayed on the SLM in order to modulate the horizontal (top) and vertical (bottom) components of the Wray field. Each circular area has a diameter of 260 pixels.
    }
    \end{figure}

 As explained in the main text, the generation of the skyrmionic textures requires the control of the amplitude and phase of the transverse polarization components. There are several experimental proposals to generate complex vector fields based on the modulation of two orthogonal polarization components. Here, we adapt the approach of Maurer {\it et al.} \cite{vector_setup} combined with the algorithm introduced by Bolduc {\it et al.} \cite{method_modulation} for encoding amplitude and phase in a phase-only hologram. 
 The setup for the generation and measurement of the skyrmionic textures is shown in Fig.~\ref{fig:experiment}(a).

 To modulate independently the two transverse polarization components of a laser beam, the SLM screen is divided into two regions. In each of these regions, a circular area is defined where the holograms are displayed. For simplicity, we modulate the horizontal and vertical components of the skyrmionic texture. These components are computed from Eqs.~\eqref{components} of the main text as 
\begin{equation}
    E_{\textbf{x},\textbf{y}}=\frac{c_{\textbf{x},\textbf{y}}}{\sqrt{2}}\left( E_\textbf{r} \pm E_\textbf{l} \right),
  \label{eq:general_field_lineal}
\end{equation}
where \textbf{x}, \textbf{y} are the horizontal and vertical lineal polarization components, $c_{\textbf{x}}=1$ and $c_{\textbf{y}}=\im$.

According to Ref.~\cite{method_modulation}, if $A$ and $\Phi$ are the desired amplitude and phase modulated by the SLM, then the displayed hologram is given by $ \Psi=M \mathrm{Mod(} F\mathrm,2\pi{)}$ (assuming an incoming beam with constant amplitude and phase), where $M=1+\frac{1}{\pi}\mathrm{sinc}^{-1}(A)$ and $F=\Phi-\pi M$. In order to perfectly recombine the two components with a Wollaston
prism (WP), a blazed phase $\Phi_{B}$ of opposite sign must be displayed in each area. In addition, a correction phase $\Phi_{C}$ (provided by the manufacturer) must be used to account for the intrinsic aberrations of the SLM, so that the final desired amplitude is given by $A_{\textbf{x},\textbf{y}}=|E_{\textbf{x},\textbf{y}}|$ and the final desired phase is $\Phi_{\textbf{x},\textbf{y}}=\mathrm{Mod}(\mathrm{arg}[E_{\textbf{x},\textbf{y}}]\pm \Phi_{B}+\Phi_{C},2\pi)$.

\begin{figure}
\centering
\def\svgwidth{0.48\textwidth}
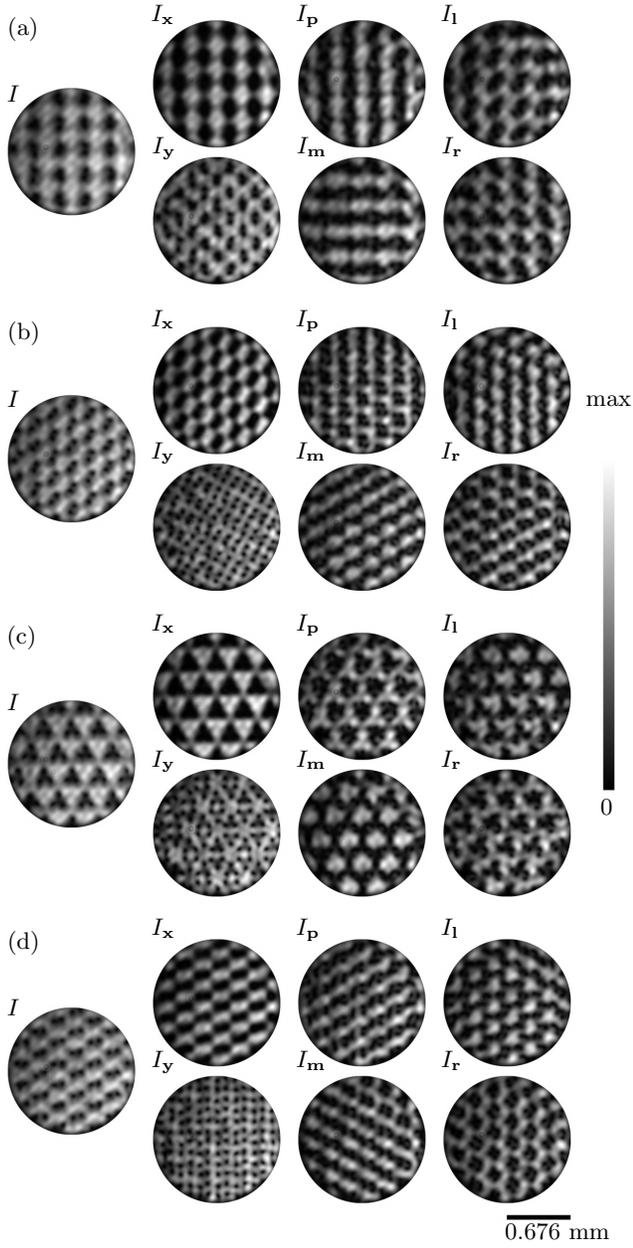\caption{\label{fig:measurements}
Measurements of the total intensity and of the intensities of each of the six polarization projections: \textbf{x}, \textbf{y}, \textbf{p}, \textbf{m}, \textbf{r} and \textbf{l} for (a) Pierce, (b) Adams, (c) Lee and (d) Wray fields. All the images are normalized with respect to their maximum. The circular area has a diameter of 261 pixels, with a pixel size of 5.20~{\textmu}m. The exposure time for each image is 2.6~s.}
\end{figure}

As shown in Fig.~\ref{fig:experiment}(a), a telescopic system projects the SLM plane onto the camera, 
and a spatial filter (SF) is used to remove spurious contributions. By rotating a linear polarizer (LP), the intensities of the components in \textbf{x}, \textbf{y}, \textbf{p} and \textbf{m} (the latter two at $\pm 45^{\circ}$ from \textbf{x}, respectively) are measured. By placing a quarter-wave plate (QWP) with fast axis at $\pm 45^{\circ}$ with respect to \textbf{x}, the intensities of the \textbf{r} and \textbf{l} components are measured. Let us refer to these intensities as $I_i$ with $i=\mathbf{x}, \mathbf{y}, \mathbf{p}, \mathbf{m}, \mathbf{l}, \mathbf{r}$, (see Fig. \ref{fig:measurements}). We also measure the total intensity $I$ by removing the LP and the QWP. We can then calculate the Stokes parameters according to:
\begin{equation}
  \begin{aligned}
   & \quad S_0 = I, \\
   & \quad S_1 = I_\textbf{x} - I_\textbf{y}, \\
   & \quad S_2 = I_\textbf{p} - I_\textbf{m},\\
   & \quad S_3 = I_\textbf{l} - I_\textbf{r}.
 \end{aligned}
\end{equation}
We compute the normalized Stokes vector as $\textbf{s}'=(S_1,S_2,S_3)/S_0$. Ideally this is a unit vector given that the field is fully polarized. However, due to noise in the measurements and the spatial extent of the pixels, the norm of this vector can be smaller (and in some cases higher) than unity. We therefore renormalize it as $\textbf{s}=\textbf{s}'/|\textbf{s}'|$.

The components of $\mathbf{s}=(s_1,s_2,s_3)$ are used to compute the experimental values of the azimuthal and polar angles on the Poincaré sphere, $\phi$ and $\theta$ respectively, as
\begin{align}
\phi&=\mathrm{atan2}(s_2,s_1),\\
\theta&=\arccos s_3.
\end{align}
The semi-major and semi-minor axes of the polarization ellipses are obtained from
\begin{align}
a&=\left[1+\frac{\tan(\theta-\pi/4)}{2}\right]^{-1/2},\\
b&=\sqrt{1-a^2}.
\end{align}

\subsection{Additional experimental results}\label{sec:additional_experiments}

\begin{figure}[h]
\centering
\def\svgwidth{0.49\textwidth}
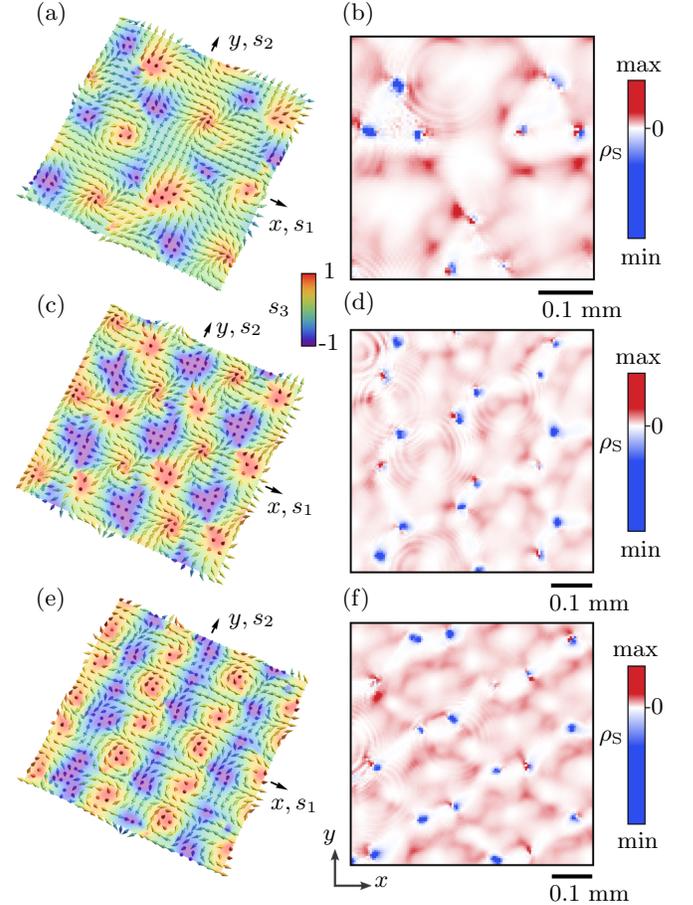\caption{Measured (a,c,e) Stokes vector distribution, and (b,d,f) Skyrme density for (a,b) Adams, (c,d) Lee and (e,f) Wray fields.\label{fig:experiments_supplemental} }
\end{figure}

\begin{figure}
\centering
\def\svgwidth{0.49\textwidth}
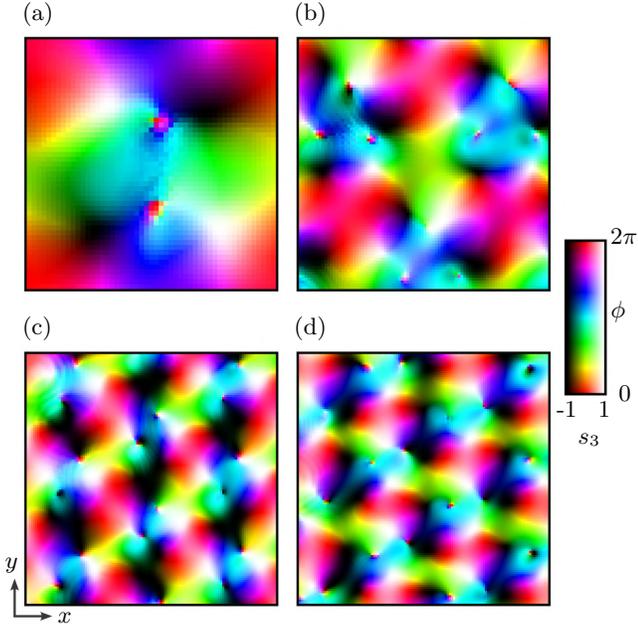\caption{Experimental $\phi, s_3$ distributions for (a) Peirce, (b) Adams, (c) Lee and (d) Wray fields.\label{fig:hue_experiment}}
\end{figure}

\begin{figure}
\centering
\def\svgwidth{0.49\textwidth}
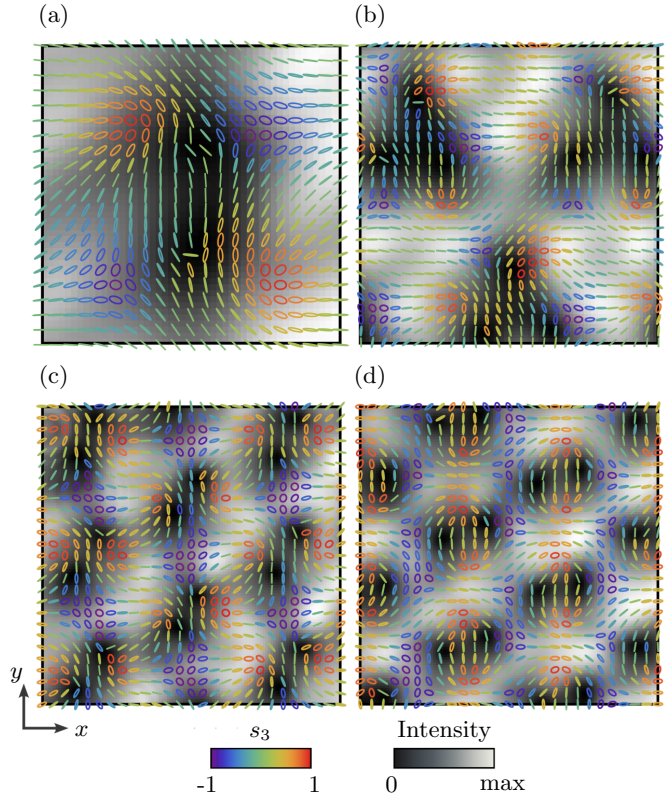\caption{Experimental polarization maps for (a) Peirce, (b) Adams, (c) Lee and (d) Wray fields.\label{fig:ellipses_experimental} }
\end{figure}

The measured Stokes vector distribution and the Skyrme density for Adams, Lee and Wray fields are shown in Fig.~\ref{fig:experiments_supplemental}. Figure \ref{fig:hue_experiment} shows the experimental distributions of $\phi$ and $s_3$ for all the fields. The measured polarization map and the intensity distribution for the Peirce, Adams, Lee, and Wray fields are depicted in Fig.~\ref{fig:ellipses_experimental}. Like for the Peirce field, the Skyrme density for the other three fields exhibits highly negative regions near points where Eqs.~\eqref{components} in the main text predict the field's zeros.

\newpage

\subsection{Instability of the zeros in the field}\label{sec:instability}
\begin{figure}[h]
\centering
\def\svgwidth{0.47\textwidth}
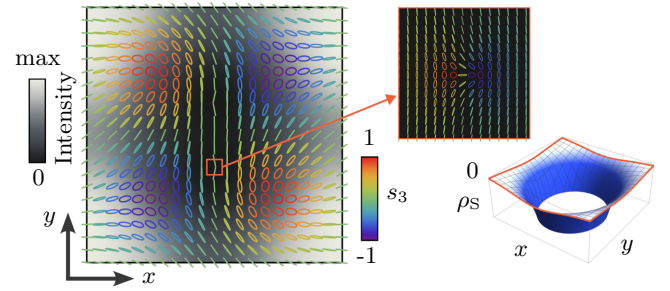\caption{\label{fig:Peirce_instability} Illustration of the impact of aperture size limitations for Peirce's field. When considering the entire spatial frequency spectrum of the field, the polarization state distribution matches the one depicted in Fig.~\ref{fig:Peirce_field_theory}(b) in the main text. However, when only the diffraction orders in Fig.~\ref{fig:spectrums}(a) are taken into account, the polarization state distribution changes to the one shown here. Near the zeros of the field in Fig.~\ref{fig:Peirce_field_theory}(b) in the main text, now a pair of C-points emanates from them. The behavior of both the polarization state and the Skyrme density near these points is illustrated here.
}
\end{figure}

\vspace*{-6.2mm}
In Fig.~\ref{fig:Peirce_experiment}(b) in the main text, we demonstrate that Peirce's optical field exhibits regions of highly negative Skyrme density, $\rho_\mathrm{S}$, at the positions where the zeros of the field are predicted to be. In fact, this occurs for all fields, as shown in Fig.~\ref{fig:experiments_supplemental}. As explained in the main text, the clipping of higher orders of the fields' spectra gives rise to $\rho_\mathrm{S}<0$ in these regions. Figure \ref{fig:Peirce_instability} shows the polarization map and the intensity of the field obtained considering only the diffraction orders we observed in the spectrum of our experimental field, which are illustrated in Fig.~\ref{fig:spectrums}(a). Note that this polarization distribution is closer to the measured polarization distribution shown in Fig.~\ref{fig:ellipses_experimental}(a), especially around the points where zeros were expected had the full spectrum been considered. It is also shown in Fig.~\ref{fig:Peirce_instability} that a right-handed lemon/left-handed star pair is generated near these points, heralding zones of highly negative Skyrme density. Right at the points where the zeros were expected, $\rho_\mathrm{S}$ can take arbitrarily large values following a small perturbation.

\end{document}